\documentclass[twocolumn]{aastex62}

\newcommand\feh{[Fe/H]}
\newcommand\alphafe{[$\alpha$/Fe]}
\newcommand\teff{$T_{\rm{eff}}$}
\newcommand\logg{$\log$ $g$}
\newcommand\paperi{E19a}
\newcommand\paperii{G19}
\newcommand\teffphot{$T_{\rm{eff,phot}}$}
\newcommand\loggphot{$\log$ $g$}
\newcommand\fehphot{[Fe/H]$_{\rm{phot}}$}
\newcommand\vhelio{$v_{\rm{helio}}$}
\newcommand\kms{km s$^{-1}$}

\graphicspath{{./}{figures/}}

\usepackage[flushleft]{threeparttable}
\usepackage{comment}

\received{2019 July 30}
\revised{2019 December 20}
\accepted{2019 December 28}
\submitjournal{ApJ}

\shorttitle{[$\alpha$/Fe] in M31's Halo, Stream, and Disk}
\shortauthors{Escala et al.}

\begin{document}

\title{Elemental Abundances in M31: A Comparative Analysis of Alpha and Iron Element Abundances in the the Outer Disk, Giant Stellar Stream, and Inner Halo of M31}

\correspondingauthor{I. Escala}
\email{iescala@caltech.edu, iescala@princeton.edu}

\author[0000-0002-9933-9551]{Ivanna Escala}
\affiliation{Department of Astronomy, California Institute of Technology, 1200 E California Blvd, Pasadena, CA, 91125, USA}
\affiliation{Department of Astrophysical Sciences, Princeton University, 4 Ivy Lane, Princeton, NJ, 08544, USA}

\author[0000-0003-0394-8377]{Karoline M. Gilbert}
\affiliation{Space Telescope Science Institute, 3700 San Martin Drive, Baltimore, MD 21218, USA}
\affiliation{Department of Physics \& Astronomy, Bloomberg Center for Physics and Astronomy, John Hopkins University, 3400 N. Charles St, Baltimore, MD 21218, USA}

\author[0000-0001-6196-5162]{Evan N. Kirby}
\affiliation{Department of Astronomy, California Institute of Technology, 1200 E California Blvd, Pasadena, CA, 91125, USA}

\author[0000-0002-3233-3032]{Jennifer Wojno}
\affiliation{Department of Physics \& Astronomy, Bloomberg Center for Physics and Astronomy, John Hopkins University, 3400 N. Charles St, Baltimore, MD 21218, USA}

\author[0000-0002-6993-0826]{Emily C. Cunningham}
\affiliation{Department of Astronomy and Astrophysics, University of California, Santa Cruz, 1156 High St, Santa Cruz, CA, 95064, USA}

\author[0000-0001-8867-4234]{Puragra Guhathakurta}
\affiliation{UCO/Lick Observatory, Department of Astronomy \& Astrophysics, University of California Santa Cruz, 
 1156 High Street, 
 Santa Cruz, California 95064, USA}



\begin{abstract}

We measured [Fe/H] and [$\alpha$/Fe] using spectral synthesis of low-resolution stellar spectroscopy for 70 individual red giant branch stars across four fields spanning the outer disk, Giant Stellar Stream (GSS), and inner halo of M31. Fields at M31-centric projected distances of 23 kpc in the halo, 12 kpc in the halo, 22 kpc in the GSS, and 26 kpc in the outer disk are $\alpha$-enhanced, with $\langle$[$\alpha$/Fe]$\rangle$ =  0.43, 0.50, 0.41, and 0.58, respectively. The 23 kpc and 12 kpc halo fields are relatively metal-poor, with $\langle$[Fe/H]$\rangle$ = $-$1.54 and $-$1.30, whereas the 22 kpc GSS and 26 kpc outer disk fields are relatively metal-rich with $\langle$[Fe/H]$\rangle$ = $-$0.84 and $-$0.92, respectively. For fields with substructure, we separated the stellar populations into kinematically hot stellar halo components and kinematically cold components. We did not find any evidence of a \added{radial} \alphafe\ gradient along the high surface brightness core of the GSS between $\sim$17$-$22 kpc. However, we found tentative suggestions of a negative \added{radial} \alphafe\ gradient in the stellar halo, which may indicate that different progenitor(s) or formation mechanisms contributed to the build up of the inner versus outer halo. Additionally, the \alphafe\ distribution of the metal-rich (\feh\ $>$ $-$1.5), smooth inner stellar halo (r$_{\rm{proj}}$ $\lesssim$ 26 kpc) is inconsistent with having formed from the disruption of progenitor(s) similar to present-day M31 satellite galaxies. 
The 26 kpc outer disk is most likely associated with the extended disk of M31, where its high $\alpha$-enhancement provides support for an episode of rapid star formation in M31's disk \added{possibly} induced by a major merger.
\end{abstract}



\section{Introduction} \label{sec:intro}
 
Stellar halos probe various stages of accretion history, as well as preserving signatures of {\it in-situ} stellar formation \citep{Zolotov2009,Cooper2010,Font2008,Font2011,Tissera2013,Tissera2014}. The stellar halo and stellar disk of $L_{\star}$ galaxies are connected through accretion events that not only build up the halo, but can impact the evolution of the disk \citep{Abadi2003,Penarrubia2006,Tissera2012}. Additionally, stellar disks can contribute to the inner stellar halo via heating mechanisms \citep{Purcell2010,McCarthy2012,Tissera2013}. The formation history of these various structural components are imprinted in its stellar populations at the time of their formation via chemical abundances \citep{Robertson2005,BullockJohnston2005,Font2006,Johnston2008,Zolotov2010,Tissera2012}. In particular, measurements of $\alpha$-element abundances (O, Ne, Mg, Si, S, Ar, Ca, and Ti) encode information concerning the relative timescales of Types Ia and \replaced{II}{core-collapse} supernovae (e.g., \citealt{GilmoreWyse1998}) and the epoch of accretion onto the host $L_{\star}$ galaxy, whereas \feh\ measurements provide information concerning the star formation duration of a stellar system. 

The Andromeda galaxy (M31) is ideal for studies of stellar halos and stellar disks, given that it is viewed nearly edge-on \citep{deVaucouleurs1958}. In contrast to the Milky Way (MW), M31 appears to be more representative of a typical spiral galaxy \citep{Hammer2007}. Thus, M31 serves as a complement to the MW in studies of galaxy formation and evolution. Although much has been learned about the global properties of M31 and its tidal debris through photometry and shallow spectroscopy (e.g., \citealt{Kalirai2006b,Ibata2005,Ibata2007,Ibata2014,Gilbert2007,Gilbert2009b,Gilbert2012,Gilbert2014,Gilbert2018,Koch2008,McConnachie2009,McConnachie2018}), the level of detail available in the MW to study its accretion history from resolved stellar populations \citep{Haywood2018,Deason2018,Helmi2018,Gallart2019,Mackereth2019a} is currently not achievable in M31.
 
In particular, the distance to M31 (785 kpc; \citealt{McConnachie2005}) has historically precluded robust spectroscopic measurements of \alphafe\ and \feh\ for individual stars. The majority of chemical information of individual RGB stars in M31 and its dwarf satellite galaxies originate from photometric metallicity estimates or spectroscopic metallicity estimates from the strength of the calcium triplet \citep{Chapman2006,Kalirai2006b,Koch2008,Kalirai2009,Richardson2009,Collins2011,Gilbert2014,Ibata2014,Ho2015}. However, the degree to which photometric and calcium triplet based metallicity estimates accurately measure iron abundance alone is uncertain \citep{Battaglia2008,Starkenburg2010,Lianou2011,DaCosta2016}. It was only in 2014 that \citeauthor{Vargas2014a}\ presented the first spectroscopic chemical abundances in the M31 system based on spectral synthesis of medium-resolution \citep{Kirby2008,Kirby2009} spectroscopy. 

Here, we present the third contribution of a deep spectroscopic survey of the stellar halo, tidal streams, disk, and present-day satellite galaxies of M31. The first work in this series (\citealt{Escala2019}, hereafter \paperi) applied a new technique of spectral synthesis of low-resolution ($R$ $\sim$ 2500) spectroscopy to individual RGB stars in the smooth, metal-poor halo of M31 at $r_{\rm{proj}}$ = 23 kpc. These were the first measurements of \alphafe\ and \feh\ of individual stars in the inner halo of M31. \replaced{Gilbert et al., submitted}{\citet{Gilbert2019}}, hereafter \paperii, presented the first \alphafe\ and \feh\ measurements in the Giant Stellar Stream (GSS; \citealt{Ibata2001}) of M31, located at $r_{\rm{proj}}$ = 17 kpc. In this work, we present \alphafe\ and \feh\ measurements for three additional fields in the inner halo at $r_{\rm{proj}}$ = 12 kpc, the GSS at $r_{\rm{proj}}$ = 22 kpc, and outer disk of M31 at $r_{\rm{proj}}$ = 26 kpc. These three fields, in addition to the smooth halo field of \paperi, all overlap with {\it Hubble Space Telescope} ({\it HST}) Advanced Camera for Surveys (ACS) pointings with inferred color-magnitude diagram based star formation histories \citep{Brown2006,Brown2007,Brown2009}. The 26 kpc outer disk field represents the first abundances in the disk of M31.

Section~\ref{sec:obs} details our observations and summarizes the properties of relevant, nearby spectroscopic fields in M31. In Section~\ref{sec:method}, we describe the changes and improvements to our abundance measurement technique (\paperi) and discuss our abundance sample selection. We define our membership criteria for M31 RGB stars and model their velocity distributions in Section~\ref{sec:kinematics}, with a focus on separating the stellar halo from substructure. Section~\ref{sec:abund} presents the full abundance distributions and separates them into kinematic components. We discuss our abundances in the context of the existing literature on M31 in Section~\ref{sec:discuss}.

\section{Observations} \label{sec:obs}

\begin{table}
\begin{threeparttable}
\caption{M31 DEIMOS Observations\tnote{a}}
\label{tab:m31_obs}
\begin{tabular*}{\columnwidth}{l @{\extracolsep{\fill}} ccccc}
\hline
\hline
Object & Date & $\theta_s$ ('') & $X$ & $t_{\textrm{exp}}$ (s) & $N$\\
\hline
\multicolumn{6}{c}{12 kpc Halo Field (H)} \\\hline
H1 & 2014 Sep 29 & 0.90 & 1.67 & 1097 & 110\\
H1 & 2014 Sep 30 & 0.90 & 2.16 & 5700 & ...\\
H1 & 2014 Oct 1 & 0.73 & 2.11 & 5700 & ...\\
H2\tnote{b} & 2014 Sep 29 & 0.9 & 1.29 & 2400 & 110\\
H2 & 2014 Sep 30 & 0.80 & 1.39 & 4200 & ...\\
H2 & 2014 Oct 1 & 0.90 & 1.32 & 4320 & ...\\ \hline
\multicolumn{6}{c}{22 kpc GSS Field (S)} \\\hline
S1 & 2014 Sep 30 & 0.70 & 1.12 & 4800 & 114\\
S1 & 2014 Oct 1 & 0.75 & 1.11 & 3600 & ...\\
S2 & 2014 Sep 29 & 0.90 & 1.07 & 2400 & 114\\
S2 & 2014 Sep 30 & 0.70 & 1.07 & 4261 & ...\\
S2 & 2014 Oct 1 & 0.75 & 1.07 & 4800 & ...\\ \hline
\multicolumn{6}{c}{26 kpc Disk Field (D)} \\\hline
D1 & 2014 Sep 30 & 0.60 & 1.15 & 4200 & 126\\
D1 & 2014 Oct 1 & 0.60 & 1.18 & 4320 & ...\\
D2 & 2014 Sep 29 & 0.70 & 1.43 & 3600 & 126\\
D2 & 2014 Sep 30 & 0.70 & 1.41 & 4683 & ...\\
D2 & 2014 Oct 1 & 0.60 & 1.41 & 4320 & ...\\
\hline
\end{tabular*}
\begin{tablenotes}
\item Note. \textemdash The columns of the table refer to slitmask name, date of observation, seeing in arcseconds, airmass, exposure time per slitmask in seconds, and number of stars targeted per slitmask.
\item[a] The observations for f130\_2, which we further analyze in this work, were published by \citet{Escala2019}.
\item[b] Slitmasks indicated ``1'' and ``2'' are identical, except that the slits on ``2'' are tilted according to the parallactic angle at the approximate time of observation.
\end{tablenotes}
\end{threeparttable}
\end{table}

\subsection{Data}

We summarize our deep M31 observations for fields H, S, and D in Table~\ref{tab:m31_obs}. The slitmasks for H, S, and D were observed for a total of 6.5, 5.5, and 5.9 hours, respectively. The M31 stars in these fields were included as additional targets on the slitmasks first presented by \citet{Cunningham2016}, which were intended to target MW foreground halo stars. We utilized the Keck/DEIMOS \citep{Faber2003} 600 line mm$^{-1}$ (600ZD) grating with the GG455 order blocking filter, a central wavelength of 7200 \AA, and 0.8'' slitwidths. Two separate slitmasks were designed for each field, with the same mask center, mask position angle, and target list, but with differing slit position angles. This enabled us to approximately track the changes in parallatic angle throughout the night, minimizing flux losses due to differential atmospheric refraction at blue wavelengths. The spectral resolution is approximately $\sim$2.8 \AA \ FWHM. As discussed in \paperi, using a low-resolution grating (comparing to the medium-resolution DEIMOS 1200G grating, $\sim$1.3 \AA \ FWHM) provides the advantage of higher signal-to-noise per pixel for the same exposure time and observing conditions. The similarly deep (5.8 hours) observations for an additional field, f130\_2, which we further analyze in this work, were published by \paperi. Additionally, we observed radial velocity templates (\S~\ref{sec:rv}; Table~\ref{tab:rv_templates}) in our science configuration. 

\subsection{Field Properties}

\begin{figure}
\includegraphics[width=\columnwidth]{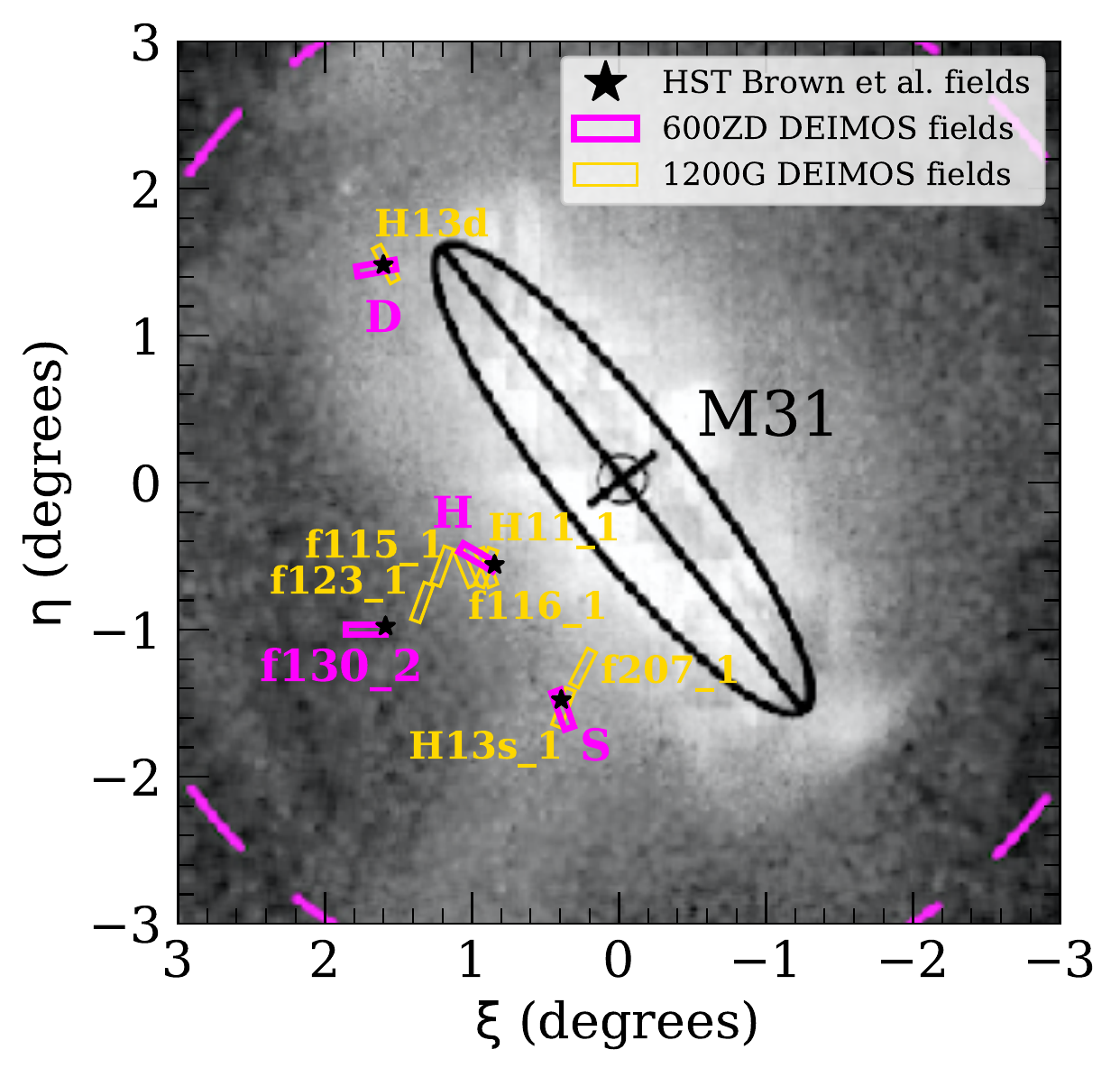}
\caption{The location of M31 DEIMOS fields observed with the 600ZD grating (\S~\ref{sec:obs} of this work, \paperi; magenta rectangles), the 1200G grating (\citealt{Kalirai2006a,Gilbert2007,Gilbert2009b}; yellow rectangles), and \textit{HST}/ACS fields (\citealt{Brown2009}; black stars) in M31-centric coordinates, overlaid on the PAndAS star count map \citep{McConnachie2018}. The dashed magenta line corresponds to 50 projected kpc. The ACS fields are represented as points given their extent on the sky (202''$\times$202'') relative to the DEIMOS masks. 
Our spectroscopic fields span M31's outer disk at 26 kpc, Giant Stellar Stream at 22 kpc, and the inner halo at 12 kpc and 23 kpc.}
\label{fig:fields}
\end{figure}

The fields H, S, and D are located at approximately 12 kpc, 22 kpc, and 26 kpc, respectively, away from the M31 galactic center in projected radius. The DEIMOS slitmasks were designed to target RGB stars near the well-studied halo21, halo11, stream, and disk fields presented in the catalog of \citet{Brown2009}. The wide \textit{HST}/ACS images were obtained in the broad $V$ and $I$ filters and reach $\sim$1.5 magnitudes fainter than the oldest main-sequence turn-off. Table~\ref{tab:m31_fields} summarizes the positioning on the sky of all four 600ZD fields and the accompanying \textit{HST}/ACS pointings. Figure~\ref{fig:fields} provides an illustration relative to the galactic center of M31 for these fields, including relevant 1200G fields (H11, H13s, H13d, and f207\_1). We also include the 1200G fields f115\_1, f116\_1, and f123 \citep{Gilbert2007} in Figure~\ref{fig:fields}, given their proximity to field H. The 1200G fields are not analyzed in this work, but their known kinematics are useful for placing our 600ZD observations in context. The dimensions of each DEIMOS slitmask are approximately 16'$\times$4', whereas the ACS images are comparatively small, spanning 202''$\times$202''.

The field S is nearly identical to H13s\_1, which was first observed for $\sim$1 hour using the 1200 line mm$^{-1}$ (1200G) grating on DEIMOS by \citet{Kalirai2006a} and later re-analyzed using an improved spectroscopic data reduction by \citet{Gilbert2009b}. Field S is located southeast of an additional Giant Stellar Stream field, f207\_1 \citep{Gilbert2009b}. 
f207\_1 is located near the eastern edge of the highest surface brightness region of the GSS core, at a projected radius of $\sim$17 kpc.  The DEIMOS 1200G fields H11 and H13d, which overlap with the southwestern and northwestern edges of the 600ZD fields H and D respectively, were also first observed by \citet{Kalirai2006a}.  Field H11 was subsequently re-analyzed by \citet{Gilbert2007} following improvements in the reduction technique. The field f130\_2, which is located at 23 kpc in projected radius, has been previously studied by \paperi, for which shallow spectroscopy was first published by \citet{Gilbert2007}.

\begin{table}
    \centering
    \begin{threeparttable}
    \caption{DEIMOS 600ZD Velocity Templates}
    \begin{tabular*}{0.8\columnwidth}{l@{\extracolsep{\fill}} ccc}
    \hline\hline
    Object & Spec. Type & X & t$_\textrm{exp}$ (s)  \\
    \hline
    HD 103095 & K1 V & 1.39 & 20 \\
    HD 122563 & G8 III & 1.37 & 20 \\
    HD 187111 & G8 III & 1.72 & 20\\  
    HD 38230 & K0 V C & 1.08 & 720\\  
    HR 4829 & A2 V C & 1.64 & 100\\
    HD 109995 & A0 V C & 1.54 & 99\\ 
    HD 151288 & K7.5 V & 1.04 & 20\\
    HD 345957 & G0 V & 1.62 & 200\\
    HD 88609 & G5 III C & 1.45 & 45\\
    HR 7346 & B9 V & 1.34 & 20\\
    \hline
    \end{tabular*}
    \label{tab:rv_templates}
    \begin{tablenotes}
    \item Note. \textemdash\ All templates were observed on 2019 Mar 10. 
    \end{tablenotes}
    \end{threeparttable}
\end{table}

\begin{table*}
\centering
\begin{threeparttable}
\caption{M31 Field Positions}
\label{tab:m31_fields}
\begin{tabular*}{1.5\columnwidth}{l @{\extracolsep{\fill}} ccclcc}
\hline
\hline
Field & $r$ (kpc)\tnote{a} & $\alpha_\textrm{J2000}$ & $\delta_\textrm{J2000}$ & P.A.\tnote{b} & ACS Field & $d_\textrm{ACS}$ (kpc)\tnote{c}\\[0.4em]
\hline
f130\_2 & 23 & 00:49:37.49 & +40:16:07.0 & +90 & halo21 & 1.44 \\
H & 12 & 00:46:34.09 & +40:45:38.6 & $-$120 & halo11 & 1.35 \\
S & 22 & 00:44:15.98 & +39:43:31.5 & +20 & stream & 0.92 \\
D & 26 & 00:49:20.59 & +42:43:44.7 & +100 & disk & 0.58 \\
\hline
\end{tabular*}
\begin{tablenotes}
\item[a] Projected radius of the mask center from M31 galactocenter.
\item[b] Slitmask position angle, in degrees east of north.
\item[c] Projected distance from DEIMOS mask center to pointing center of corresponding ACS field.
\end{tablenotes}
\end{threeparttable}
\end{table*}

Based on the nearby 1200G fields, we expect that the properties of fields H, S, and D will generally reflect the inner halo of M31, the GSS, and the outer northeastern disk of M31, respectively, although other components are present in these fields. In particular, field H is likely polluted by stars belonging to a substructure known as the Southeast shelf, which is associated with the GSS progenitor (\S~\ref{sec:se_shelf}; \citealt{Gilbert2007,Fardal2007}). Field S should contain a secondary kinematically cold component of unknown origin in addition to the GSS core \citep{Kalirai2006a,Gilbert2009b,Gilbert2019}. \paperi\ showed that f130\_2 is likely associated with the ``smooth'', metal-poor component of M31's stellar halo. We refer to fields H, S, D, and f130\_2 interchangeably as the 12 kpc inner halo, 22 kpc GSS, 26 kpc outer disk, and 23 kpc smooth halo fields where appropriate to emphasize the physical properties of the M31 fields.

\section{Abundance Determination} \label{sec:method}

We use spectral synthesis of low-resolution stellar spectroscopy  (\paperi) to measure stellar parameters and abundances from our deep observations of M31 RGB stars. In summary, we measure \feh \ and \alphafe \ from regions of the spectrum sensitive to Fe and $\alpha$-elements (Mg, Si, Ca), respectively, by comparing to a grid of synthetic spectra degraded to the resolution of the DEIMOS 600ZD grating. We also measure the spectroscopic effective temperature, \teff, informed by photometric constraints, and fix the surface gravity, \logg, to the photometric value. \added{Measurements of \feh\ and \alphafe\ using spectra obtained with the 600ZD grating are generally consistent with equivalent measurements \citep{Kirby2008} from 1200G spectra (Appendix~\ref{sec:600_vs_1200}).} For a detailed description of the \added{low-resolution spectral synthesis} method, see \paperi. In the following subsections, we describe improvements and changes to our technique since \paperi.

\subsection{Photometry}
\label{sec:phot}

\begin{figure*}
\centering
\includegraphics[width=0.9\textwidth]{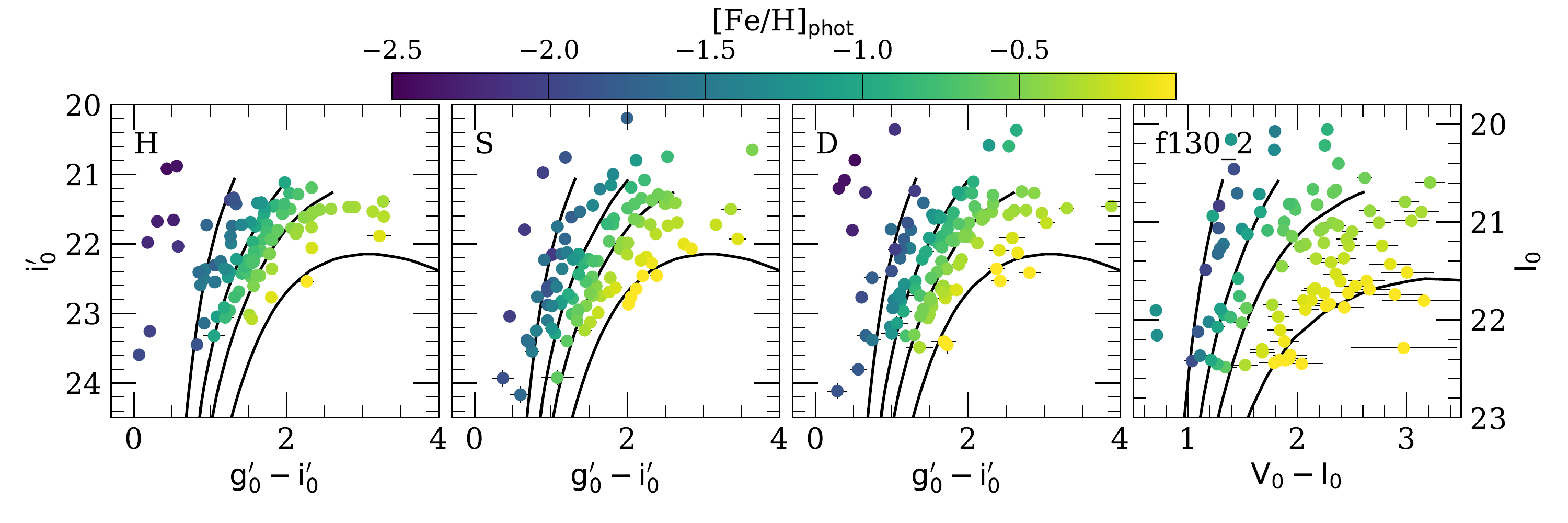}
\caption{($i_0^\prime$, $g_0^\prime-i_0^\prime$) and ($V_0-I_0$, $I_0$) color-magnitude diagrams for all stars (M31 RGB stars and MW foreground dwarf stars) in the 12 kpc inner halo field (H), 22 kpc GSS field (S), the 26 kpc outer disk field (D), and the 23 kpc smooth halo field (f130\_2). The points are color coded according to the photometric metallicity estimated for each star from the PARSEC \citep{Marigo2017} isochrones ($-2.2$ $<$ [Fe/H] $<$ $+0.2$) assuming an age of 9 Gyr for H, S, and D and 12 Gyr for f130\_2 (\S~\ref{sec:phot}). For reference, we overplot a few PARSEC isochrones (\textit{black lines}) with, from left to right, [Fe/H] = $-2.2$, $-1.1$, $-0.5$, and $+0.1$. Stars that are bluer than the most metal-poor isochrone (taking into account photometric errors) are likely MW dwarf stars.}
\label{fig:photmetal}
\end{figure*}

We utilized wide-field (1 deg$^2$) $g^\prime$ and $i^\prime$ band photometry
from the Pan-Andromeda Archaeological Survey (PAndAS) catalog \citep{McConnachie2018} for the fields H, S, and D. The images were obtained from MegaCam on the 3.6 m Canada-France-Hawaii Telescope. We extinction-corrected the photometry assuming field-specific interstellar reddening values
from the dust reddening maps of \citet{Schlegel1998}, with the corrections defined by \citet{SchlaflyFinkbeiner2011}. We used the conversion between reddening and extinction adopted by \citet{Ibata2014}. For stars present in the DEIMOS fields but absent from the PAndAS point source catalog ($\sim$20-30\% of M31 RGB stars on a given slitmask), we sourced photometry from CFHT/MegaCam images obtained by \citet{Kalirai2006a} and reduced with the CFHT MegaPipe pipeline \citep{Gwyn2008}. We cross-validated the MegaPipe photometry against that of PAndAS for common stars to verify \deleted{the} that the photometry is accurate for the majority of stars.

\begin{figure*}
    \centering
    \includegraphics[width=0.9\textwidth]{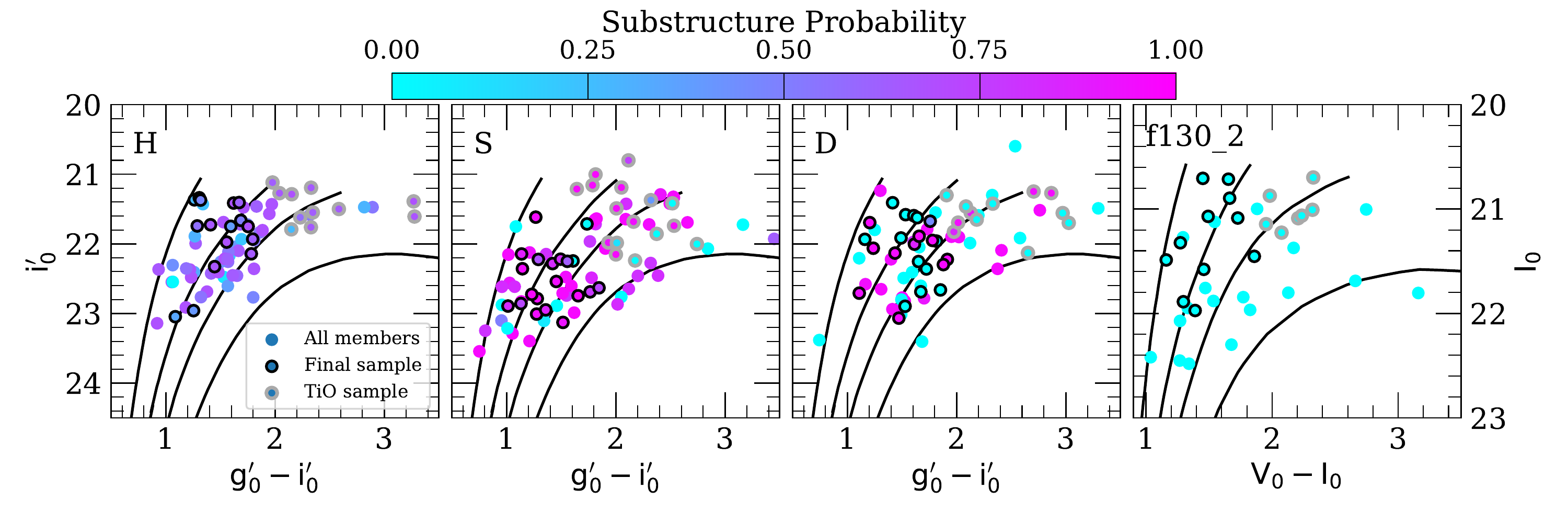}
    \caption{Color magnitude diagrams of M31 RGB stars (\S~\ref{sec:membership}) reflecting selection effects in the 12 kpc inner halo field (H), 22 kpc GSS field (S), 26 kpc outer disk field (D), and 23 kpc smooth halo field (f130\_2). Stars are color-coded according to probability of belonging to \textit{any} substructure component in a given field (\S~\ref{sec:prob_substruct}). Magenta (blue) points are likely (unlikely) to be associated with substructure. For each field, we show all M31 RGB stars, M31 RGB stars with spectroscopic abundance measurements constituting our final sample of 70 total stars (\textit{black outlined circles}; \S~\ref{sec:sample}), and M31 RGB stars with spectroscopic abundance measurements that otherwise pass our selection criteria, but show signatures of TiO absorption (\textit{grey outlined circles}). We overplot PARSEC isochrones in the appropriate filter for each field with, from left to right, [Fe/H] = $-$2.2, $-$1.0, $-$0.5, and 0. 
    On average, the omission of TiO stars from our final sample results in a bias against red stars ($g^\prime_0 - i^\prime_0 \gtrsim 2$ and $V_0 - I_0 \gtrsim 2$), which disproportionately affects the substructure components (relative to the stellar halo components) in the 12 kpc halo field and 22 kpc GSS field.}
    \label{fig:m31_cmd}
\end{figure*}

In contrast to \paperi, we did not use multiple isochrone sets to calculate photometrically-based quantities such as \teffphot \ and \logg. We employed the most recent version of the PARSEC \citep{Marigo2017} isochrones, which are available in the relevant filters for a wide range of stellar ages and metallicities between $-2.2$ $<$ \feh\ $<$ 0.2 and \alphafe\ = 0. For stars positioned above the tip of the red giant branch\footnote{Stars that have magnitudes brighter than the tip of the red giant branch, according to the assumed isochrone set and distance modulus, are either a consequence of photometric errors or AGB stars. None of these stars are in our final abundance sample (Figure~\ref{fig:m31_cmd}).}, we linearly extrapolate to obtain estimates of \teffphot, \logg, and \fehphot. Similarly, we extrapolate blueward of the most metal-poor isochrone to determine \teffphot \ and \logg \ for these stars. We assumed a distance modulus relative to M31 of $m-M$ = 24.63 $\pm$ 0.20 \citep{Clementini2011}. We utilized the same Johnson-Cousins $V,I$ photometry for f130\_2 as in \paperi, but determined photometric parameters using the PARSEC isochrones as described above. Figure~\ref{fig:photmetal} illustrates our usage of the PARSEC isochrones to determine photometrically-based quantities, where we have color-coded the color-magnitude diagrams (CMDs) according to the estimated photometric metallicity. We assumed ages of 9 Gyr for H, S, and D based on mean stellar ages of 9.7 Gyr, 8.8 Gyr, and 7.5 Gyr, respectively, in the corresponding ACS fields (Table~\ref{tab:m31_fields}) inferred from CMD-based star formation histories \citep{Brown2006}. For f130\_2, we assumed an age of 12 Gyr, where it was inferred to have a mean stellar age of 11 Gyr \citep{Brown2007}. 

Although other isochrone sets (e.g., \citealt{Dotter2007,Dotter2008}) are also available in these filters, we based our selection in part on whether the isochrones contained contributions from molecular TiO (\S~\ref{sec:sample}) in the stellar atmosphere models used to compute the evolutionary tracks. 

\subsection{Spectral Resolution}

Previously, we approximated the spectral resolution as constant with respect to wavelength ($\Delta \lambda$). In \paperi, we determined $\Delta \lambda$ for each star by fitting the observed spectrum in a narrow range centered on the expected resolution for the 600ZD grating. In actuality, for our DEIMOS configuration, the spectral resolution is a slowly varying function of wavelength. 

We employed this approximation to circumvent the problem of an insufficient number of sky lines at bluer wavelengths to empirically determine the spectral resolution as a function of wavelength ($\Delta \lambda (\lambda)$). Alternatively, including arc lines in the fitting procedure can provide constraints in this wavelength regime. Using a combination of Gaussian widths from both sky lines and arc lines, we utilized a maximum-likelihood approach (McKinnon et al., in preparation) to determine $\Delta \lambda (\lambda)$ for each star. In the few cases per slitmask where the spectral resolution determination fails (e.g., owing to an insufficient number of arc and sky lines), we assumed $\Delta \lambda$ = 2.8 \AA, the expected resolution of the 600ZD grating (\paperi). For the case of multiple observations per star, we calculate $\Delta \lambda (\lambda)$ as the average of the individual measurements on different dates of observation for a given star.

In addition to $\Delta \lambda (\lambda)$, we determined a resolution scale parameter. This parameter accounts for the fact that the resolution as calculated from the sky lines and arc lines, which fill the entire slit, slightly overestimates $\Delta \lambda$ for the stellar spectrum, whose width depends on seeing. First, we included the resolution scale as a free parameter in our abundance determination, measuring its value, $f_i$, for each object on a given slitmask. However, given that each individual measurement is subject to noise, the final measurement, $f$, is the average of the individual measurements for the entire slitmask. The resolution scale parameter is primarily a function of seeing, and therefore should be constant for a single slitmask. In the final abundance determination, we fixed the spectral resolution at $f \Delta \lambda (\lambda)$.

Based on our globular cluster calibration sample from \paperi, we confirmed that utilizing $\Delta \lambda (\lambda)$, as opposed to the $\Delta \lambda$ approximation, alters our abundances within the 1$\sigma$ uncertainties. We re-calculated the systematic error in \feh\ and \alphafe\ from the internal spread in globular clusters, finding $\delta$([Fe/H])$_{\rm{sys}}$ = \textcolor{red}{0.130} and $\delta$([$\alpha$/Fe])$_{\rm{sys}}$ = 0.107. We repeated our chemical abundance analysis for f130\_2 (\paperi), fixing $\Delta \lambda (\lambda)$ to its empirically derived value for each star, and present these abundances in \S~\ref{sec:abund}.

\subsection{Radial Velocity}
\label{sec:rv}


We cross-correlated the observed spectrum with empirical templates of high signal-to-noise (S/N) stars \citep{Cooper2012,Newman2013}, which we observed with the 600ZD grating 
in our science configuration (Table~\ref{tab:rv_templates}). We shifted the templates to the rest frame based on their {\it Gaia} DR2 \citep{GaiaCollaboration2016, GaiaCollaboration2018b} radial velocities, except for HD 109995 \citep{Gontcharov2006}. The templates do not possess any A-band velocity offsets, as the template stars were trailed through the slit while observing.
We utilized the full template spectrum ($\sim4500-11000$ \AA) to shift the science spectrum into the rest frame. In cases where the full-spectrum radial velocity determination failed, we instead utilized the wavelength regions near the calcium triplet (8450 \AA \ $<$ $\lambda$ $<$ 8700 \AA). Additionally, we apply an A-band correction, which significantly impacts the determination of the heliocentric velocity. We determined random velocity errors from Monte Carlo resampling with 10$^{3}$ trials.

Following an improvement in the spectroscopic data reduction, \citet{Gilbert2009b} and \citet{Gilbert2018} found a typical velocity precision of $\sim5-7$ km s$^{-1}$ for low S/N ($\sim$ 10$-$12 \AA$^{-1}$) M31 RGB stars observed with the 1200G grating, including a systematic component of $\sim2$ km s$^{-1}$ from repeat observations of stars \citep{SimonGeha2007}. For our entire sample (including MW dwarf stars) with successful radial velocity measurements, our median velocity uncertainty is 11.6 km s$^{-1}$, incorporating a systematic error term for the 600ZD grating based on repeat observations of over 300 stars (5.6 \kms; \citealt{Collins2011}). The reduced velocity precision for the 600ZD grating is a consequence of its lower spectral resolution.

\subsection{Abundance Sample Selection}
\label{sec:sample}

\begin{figure*}
    \centering
    \includegraphics[width=0.9\textwidth]{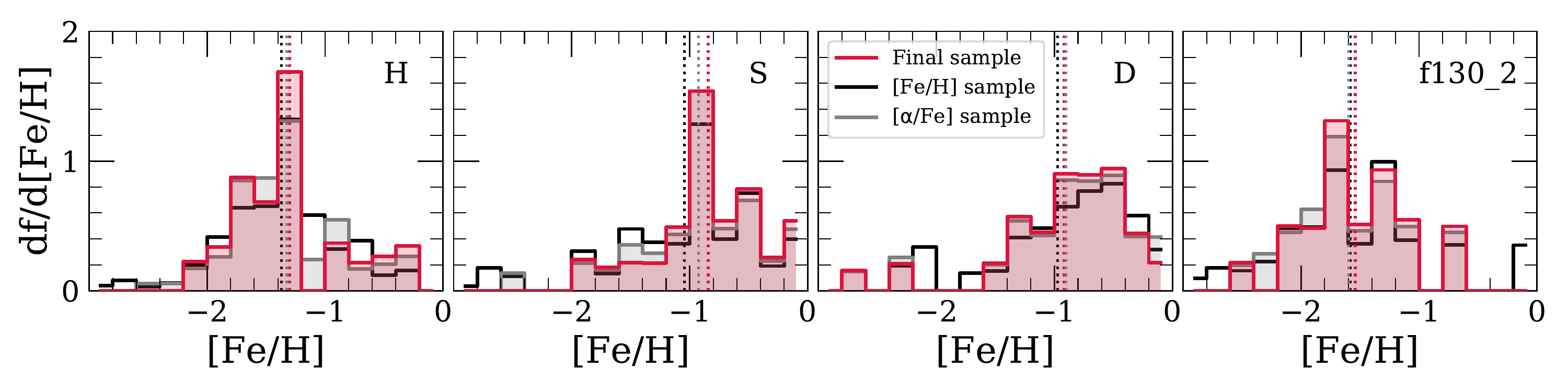}
    \caption{Metallicity distribution functions (MDFs), represented in terms of probability density, for subsets of our total sample of M31 RGB stars with abundance measurements (\S~\ref{sec:sample_select}). The MDFs are weighted according to the inverse variance of the total measurement uncertainty in [Fe/H]. We have subdivided our total sample into M31 RGB stars with a [Fe/H] measurement (\textit{black histogram}) and with both a [Fe/H] and [$\alpha$/Fe] measurement (\textit{grey histogram}), regardless of the error on the measurement. The red histogram shows our final sample, which includes M31 RGB stars with both [Fe/H] and [$\alpha$/Fe] measured to within 0.5 dex. Stars with TiO absorption are excluded from all subsets. The inverse-variance weighted means of all samples are indicated as dashed verticle lines. The metallicity distributions of the finalized sample are weakly biased against metal-poor stars with low S/N spectra.}
    \label{fig:m31_mdfs}
\end{figure*}

As in \paperi, we included only \textit{reliable} measurements for M31 RGB stars (\S~\ref{sec:membership}) in our final samples, i.e., $\delta$\feh \ $<$ 0.5,  $\delta$\alphafe \ $<$ 0.5, and well-constrained parameter estimates based on the 5$\sigma$ $\chi^2$ contours for all fitted parameters. Unreliable abundance measurements are often a consequence of insufficient S/N. In addition, we excluded spectra of stars with sufficient S/N for a reliable measurement that found a minimum at the cool end of the \teff\ range (3500 K) spanned by our grid of synthetic spectra. We also manually screened member stars for evidence of strong molecular TiO absorption between 7055$-$7245 \AA, finding that 41\%, 44\%, 34\%, and 39\% of the measurements for H, S, D, and f130\_2 passing the reliability cuts were affected by TiO. We excluded these stars from the subsequent abundance analysis, given our uncertainty in our ability to accurately and precisely measure abundances for stars with TiO in the absence of a suitable calibration sample. We found that 16, 20, 23, and 11 of the measurements in H, S, D, and f130\_2 can be considered reliable based on the above criteria, resulting in a final sample of 70 stars. Our sample of stars affected by TiO that otherwise pass our selection criteria is composed of 46 stars across all four fields.

\subsection{Selection Effects on the Abundance Distributions}
\label{sec:sample_select}

Given that our spectroscopic abundance determination is S/N limited and it is unclear how the omission of TiO from our linelist (\paperi) impacts our abundance measurements for stars with strong TiO absorption, we investigated the impact of our selection criteria  (\S~\ref{sec:sample}) on the properties of our final sample. Figure~\ref{fig:m31_cmd} shows CMDs of all M31 RGB stars (\S~\ref{sec:membership}) in each field, where we have highlighted our final sample. We also show the subset of stars with spectroscopic evidence of TiO absorption that otherwise pass our selection criteria. Excluding stars with TiO translates to an effective color bias of $g^\prime_0 - i^\prime_0 \lesssim 2$ and $V_0 - I_0 \lesssim 2$. Additionally, our final sample probes brighter magnitudes, particularly in H. Fainter stars tend to have lower S/N, which results in either an uncertain (e.g., $\delta$(\alphafe) $>$ 0.5) or failed abundance measurement. In principle, this should not affect the metallicity distribution, so long as the final sample spans the majority of the color range of the CMD. 


\begin{figure*}
\includegraphics[width=\textwidth]{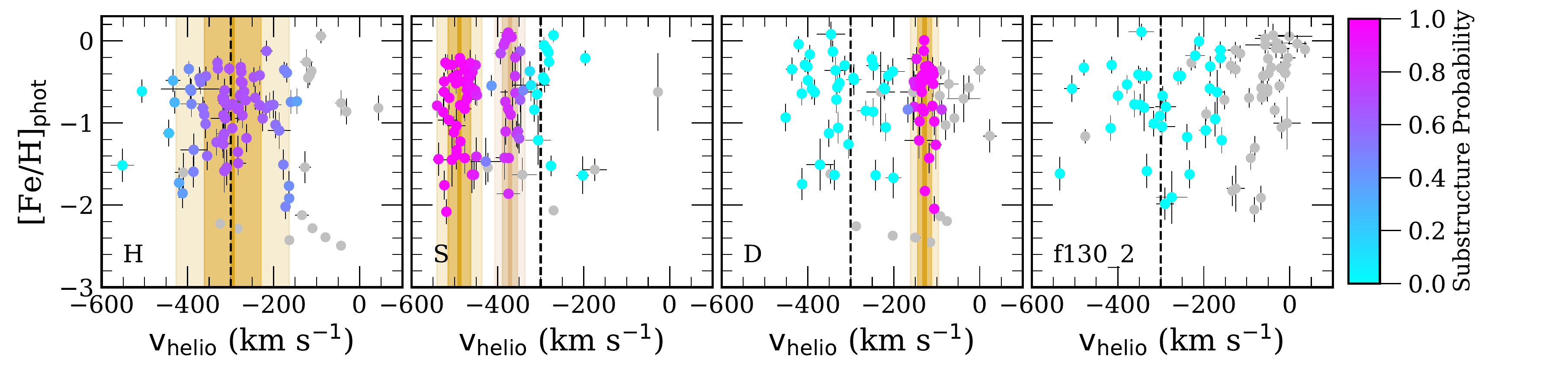}
\caption{Heliocentric radial velocity versus photometric metallicity (\S~\ref{sec:phot}) for stars with successful velocity measurements in the 12 kpc halo (H), 22 kpc GSS (S), 26 kpc disk (D), and 23 kpc halo (f130\_2) M31 fields. The velocity errors represent only the random component of the uncertainty. M31's systemic velocity ($v_{\rm{M31}} = -300$ \kms) is indicated as a dashed vertical line. Stars classified as MW foreground dwarfs (\S~\ref{sec:membership}) are shown in grey, whereas stars classified as M31 RGB stars are color-coded according to its probability of belonging to substructure, as in Figure~\ref{fig:m31_cmd}. The vertical bands represent the mean velocity and the velocity dispersion ($\mu$, $\mu\pm\sigma$, $\mu\pm2\sigma$; Table~\ref{tab:kinematic_decomp}) of the primary (\textit{orange}) and secondary (\textit{tan}) substructure components in each field. We identified 73, 84, 68, and 36 stars as M31 RGB stars in the 12 kpc, 22 kpc, 26 kpc, and 23 kpc fields, respectively.} 
\label{fig:vhelio_vs_fehphot}
\end{figure*}

We quantitatively assessed the metallicity bias introduced by excluding M31 RGB stars with imprecise spectroscopic abundance measurements. We separated our sample of M31 RGB stars with spectroscopic [Fe/H] measurements into three subsets: (1) a sample with successful [Fe/H] determinations as dictated by the 5$\sigma$ $\chi^2$ contours, with no restrictions on the errors, (2) a sample with both successful \feh\ and \alphafe\ measurements, with no restrictions on the errors, and (3) our final sample, with successful \feh\ and \alphafe\ measurements and $\delta$(\feh) $<$ 0.5 and $\delta$(\alphafe) $<$ 0.5. All three subsets exclude TiO stars. As illustrated by Figure~\ref{fig:m31_mdfs}, the inverse-variance weighted metallicity distribution functions appear similar between the three subsets. The error-weighted mean metallicity for the most inclusive sample is more metal-poor than the final sample by $\sim$0.04$-$0.07 dex for fields H, D, and f130\_2. The difference between samples is 0.20 dex for field S, owing to very metal poor stars present in sample (1) that were omitted from sample (3). 
If we assume that sample (1) better represents the true spectroscopic metallicity of M31 RGB stars in the field, then we can conclude that S/N limitations, which increase measurement uncertainty, results in a weak bias in our final sample against metal-poor stars with low S/N spectra.

Regarding the known color bias introduced by excluding TiO stars, we analyzed the photometric metallicity distributions of each sample. The isochrone set we employed to calculate \teff, \logg, and \fehphot\ (\S~\ref{sec:phot}) for H, S, and D (PARSEC; \citealt{Marigo2017}) were generated using models that included molecular TiO absorption. This allowed us to estimate the metallicity of all M31 RGB stars, many of which are not included in our final sample. Assuming [$\alpha$/Fe] = 0, we found that our final sample is biased toward lower \fehphot\ relative to the full sample of M31 RGB stars. We found that $\langle$[Fe/H]$\rangle_{\rm{phot}}$ = $-$0.89, $-$0.76, $-$0.69, and $-$0.76 for all M31 RGB stars in H, S, D, and f130\_2, respectively. For our final sample, we found that $\langle$[Fe/H]$\rangle_{\rm{phot}}$ = $-$1.17, $-$0.96, $-$0.87, and $-$1.15 for H, S, D, and f130\_2. Thus, on average, our final sample is biased toward lower \fehphot\ by $\sim$0.2$-$0.4 dex. Much of this effect is a consequence of the exclusion of TiO stars from the final sample. Including the subset of TiO stars, we obtain $\langle$[Fe/H]$\rangle_{\rm{phot}}$ = $-$0.91 dex, $-$0.85 dex, $-$0.73 dex, and $-$0.86 dex, for H, S,  D, and f130\_2, reducing the bias in the final sample to $\sim$0.02$-$0.10 dex more metal poor than the full sample. Based on this, we can conclude the primary source of bias against metal-rich stars originates from excluding TiO stars. 
However, the exact amount by which we might be biased in \feh\ is unclear, given that \fehphot, which has no knowledge of \alphafe\ and is degenerate with stellar age, cannot be translated into spectroscopic \feh. 

We do not anticipate that selection effects impacting the color distribution of our final sample incur a bias in \alphafe\ relative to the full sample of M31 RGB stars. The width, or color range, of the RGB is largely dictated by [Fe/H], as opposed to $\alpha$-enhancement \citep{Gallart2005}. However, S/N limitations may  affect the \alphafe\ distribution of the final sample, resulting in a weak bias against $\alpha$-poor stars with low S/N spectra.\footnote{Additionally, if our abundance measurements of TiO stars are indeed valid, we cannot eliminate the possibility that our final sample is biased toward lower \alphafe\ by $\sim$ 0.1$-$0.2 dex (e.g., Figure~\ref{fig:m31_abund}).}

\section{Kinematic Analysis of the M31 Fields}
\label{sec:kinematics}

\subsection{M31 Membership} \label{sec:membership}

\begin{figure*}
\centering
\includegraphics[width=0.85\textwidth]{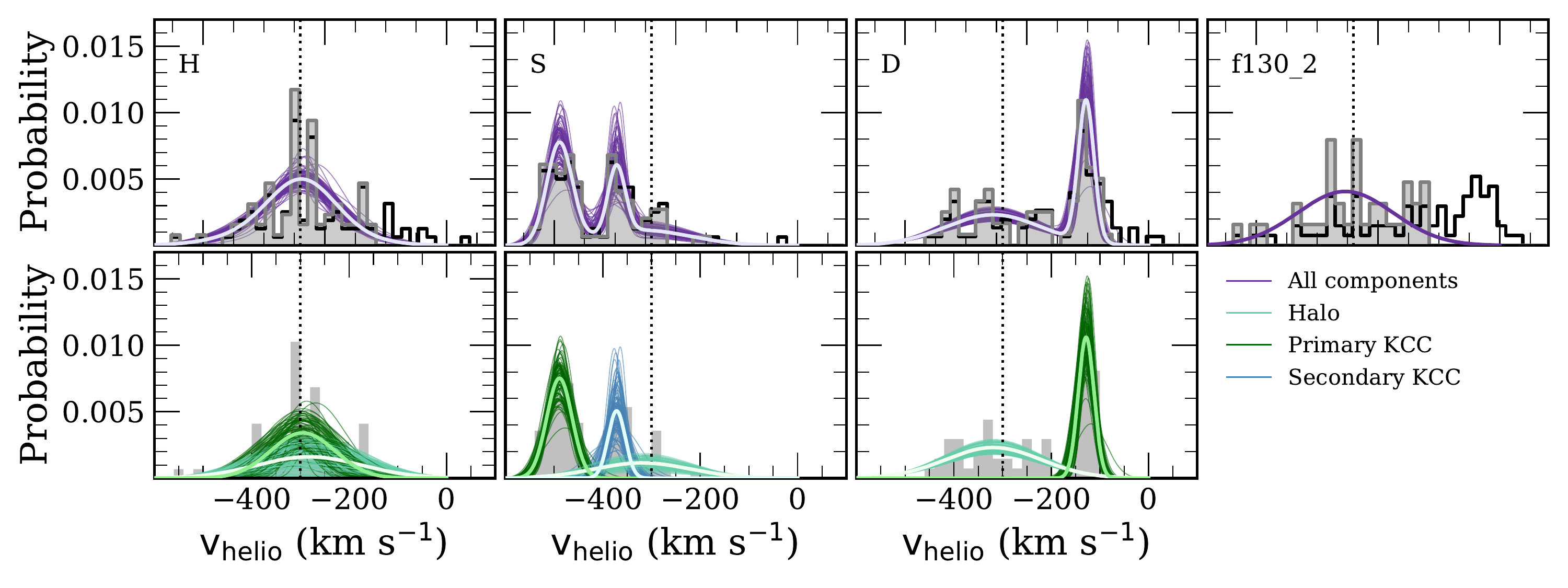}
\caption{Heliocentric radial velocity distributions of stars with successful velocity measurements (\S~\ref{sec:rv}, \textit{top panels, black histograms}), including foreground MW dwarf stars (\S~\ref{sec:membership}), and velocity distributions for M31 RGB stars (\textit{grey filled histograms}) in the 12 kpc (H), 22 kpc (S), 26 kpc (D), and 23 kpc (f130\_2) fields. We also show full velocity distribution models for M31 RGB stars (\textit{top panels, purple lines}) and, for fields with substructure, models of the kinematic components (\textit{bottom panels}). We include both stellar halo  (\textit{light green lines}) and kinematically cold components (KCCs; \textit{dark green and blue lines}) (\S~\ref{sec:kinematic_decomp}).
M31's systemtic velocity is indicated as a dotted vertical line. For fields with substructure, we show 100 randomly sampled models from the converged portion of the MCMC chain to represent the uncertainty in fitting the velocity distribution. For each field, we also show the full velocity model (and its components, where applicable) as defined by the 50$^{\rm{th}}$ percentile parameter values (\textit{thick lines}; Table~\ref{tab:kinematic_decomp}). All fields show evidence for M31 halo stars (distributed in a kinematically hot component around the systemic velocity). The GSS is located in the 22 kpc field at approximately $-$490 km s$^{-1}$, including the KCC of unknown origin at $-$370 km s$^{-1}$ \citep{Kalirai2006a,Gilbert2009b}. The 12 kpc substructure likely corresponds to the Southeast shelf (\S~\ref{sec:se_shelf}; \citealt{Fardal2007,Gilbert2007}), a tidal feature originating from the GSS progenitor. M31's disk appears as the prominent feature centered at $-$130 km s$^{-1}$ in the 26 kpc field.}
\label{fig:kinematic_decomp}
\end{figure*}

Given that foreground MW dwarf stars and M31 stars are spatially coincident and exhibit significant overlap in both their velocity distributions and CMDs, identifying bona fide M31 RGB stars is nontrivial. In \paperi, we utilized the probabilistic method of \citet{Gilbert2006} to carefully assess the likelihood of membership for stars in our spectroscopic sample. This method incorporates up to four criteria to determine membership for a majority of M31 fields: the strength of the \ion{Na}{1} $\lambda\lambda$8190 absorption line doublet, the \textit{(V, I)} color-magnitude diagram location, photometric versus spectroscopic (\ion{Ca}{2} $\lambda\lambda$8500) metallicity estimates, and the heliocentric radial velocity. However, we cannot use this exact approach for fields H, S, and D, owing to the diversity of utilized photometric filters.

Instead, we determine membership based on three criteria: (1) \ion{Na}{1} $\lambda\lambda$8190 absorption strength, (2) CMD location, and (3) heliocentric radial velocity. Given that the strength of the \ion{Na}{1} doublet depends on surface gravity, it can effectively separate M31 RGB stars from foreground MW M dwarfs \citep{Schiavon1997}. We excluded stars with clear signatures of the \ion{Na}{1} doublet as nonmembers of M31. We classified stars as MW dwarf stars if they have colors bluer than the most metal-poor isochrone (\S~\ref{sec:phot}) by an amount greater than their photometric error. Such stars are $\gtrsim$ 10 times more likely to be MW dwarf stars than M31 RGB stars \citep{Gilbert2006}. Lastly, we adopted a radial velocity cut of $v_{\rm{helio}}$ $< -150$ km s$^{-1}$ for fields H, S, and f130\_2 to select for M31 RGB stars. Using a sample of $\gtrsim$ 1000 probablistically identified M31 RGB stars, \citet{Gilbert2007} found that contamination from MW dwarf stars is largely constrained to  $v_{\rm{helio}}$ $> -150$ km s$^{-1}$. The estimated contamination fraction using this radial velocity cut, in combination with the additional membership diagnostics of \citet{Gilbert2006}, is 2-5\% across their entire sample, where contamination is defined as the fraction of bona fide MW dwarf stars classified as M31 RGB stars.

To evaluate the performance of our binary membership determination, we compared our results to those of stars with \citet{Gilbert2006} membership probabilities. For fields H, S, and f130\_2, 11\%, 24\%, and 74\%, respectively, of our sample with successful radial velocity measurements have associated M31 membership probabilities. Assuming $m-M$ = 24.47 mag \citep{Gilbert2006}, we accurately recovered 87\%, 98\%, and 97\% of both secure and marginal M31 members, including radial velocity as a membership diagnostic ($L_i > 0$; \citealt{Gilbert2012}), in H, S, and f130\_2, respectively. The fraction of stars present in our M31 RGB samples that are classified as MW dwarf stars using the method of \citet{Gilbert2006} is 0\% across all three fields. Given that we used similar membership criteria to \citet{Gilbert2006} and were able to reproduce their results to high confidence, we estimate that our true MW contamination fraction is $\sim$2-5\% across fields H, S, and f130\_2.

\begin{table*}
\centering
\begin{threeparttable}
\caption{Velocity Distribution Model Parameters for M31 Fields}
\label{tab:kinematic_decomp}
\begin{tabular*}{\textwidth}{lccccccccc}
\hline
\hline
Field & \multicolumn{1}{p{1.5cm}}{\centering $r$\\(kpc)}
& \multicolumn{1}{p{1.5cm}}{\centering $\mu_{\rm{halo}}$\\(\kms)}
& \multicolumn{1}{p{1.5cm}}{\centering $\sigma_{\rm{halo}}$\\(\kms)}
& \multicolumn{1}{p{1.5cm}}{\centering $\mu_{\rm{kcc1}}$\\(\kms)}
& \multicolumn{1}{p{1.5cm}}{\centering $\sigma_{\rm{kcc1}}$\\(\kms)}
& $f_1$ 
& \multicolumn{1}{p{1.5cm}}{\centering  $\mu_{\rm{kcc2}}$\\(\kms)}
& \multicolumn{1}{p{1.5cm}}{\centering  $\sigma_{\rm{kcc2}}$\\(\kms)}
& $f_2$\\[0.4em]
\hline
H & 12 & $-$315  & 108 & $-$295 $\pm$ 12 & 66$^{+11}_{-16}$ & 0.56$^{+0.23}_{-0.25}$ & \nodata & \nodata & \nodata \\
S & 22 & $-$319 & 98 & $-$489 $\pm$ 4 & 26 $\pm$ 3 & 0.49 $\pm$ 0.06 & $-$372 $\pm$ 5 & 17$^{+7}_{-4}$ & 0.22$^{+0.07}_{-0.06}$\\
D & 26 & $-$319 & 98 & $-$128 $\pm$ 3 & 16$^{+3}_{-2}$ & 0.43 $\pm$ 0.06 & \nodata & \nodata & \nodata \\
f130\_2 & 23 & $-$317 & 98 & \nodata & \nodata & \nodata & \nodata & \nodata & \nodata\\
\hline
\end{tabular*}
\begin{tablenotes}
\item Note. \textemdash\ The parameters describing the model components are mean velocity ($\mu$), velocity dispersion ($\sigma$), and normalized fractional contribution ($f$), where components are separated into the kinematically hot stellar halo and kinematically cold components (KCCs). Mean parameter values are expressed as the 50$^{\rm{th}}$ percentile values of the corresponding marginalized posterior probability distributions. The errors on each parameter are calculated based on the 16$^{\rm{th}}$ and 84$^{\rm{th}}$ percentiles. Parameters for the halo components are adopted from \citet{Gilbert2018}. In the case of the 22 kpc GSS field, the primary KCC is the GSS core.
\end{tablenotes}
\end{threeparttable}
\end{table*}

Stars in field D do not possess previously determined membership probabilities to which we could compare. The rotation of M31's northeastern disk produces a redshift relative to M31's systemic velocity, such that the peak of the disk is located at $v_{\rm{disk}}$ $\sim$ $-$130 \kms\ (\S~\ref{sec:kinematics_results}). The presence of the disk invalidates the use of a \vhelio\ $<$ $-$150 \kms\ velocity cut as a diagnostic for M31 membership in field D. Instead, we employed a less conservative radial velocity cut of $v_{\rm{helio}} < -100$ \kms\ to identify potential M31 RGB stars. This cut likely recovers the majority of M31 members in this field, but increases the MW contamination fraction (in the velocity range of $-150$ \kms\ $< v_{\rm{helio}} < -100$ \kms). \citet{Gilbert2006} estimated that \textit{only} using a radial velocity cut of $v_{\rm{helio}} < -100$ \kms\ results in a 10\% contamination fraction in inner halo fields. For disk fields covering an area of 240 arcmin$^{2}$ within their color-magnitude selection window, including outer disk fields in the northeastern quadrant, \citet{Ibata2005} used predictions from Galactic models \citep{Robin2003} to argue that MW contamination in disk fields is negligible ($\sim$5\%) for  $v_{\rm{helio}} < -100$ \kms. Therefore, we expect that the MW contamination fraction in field D is $\sim$5-10\%, where the relatively high density of stars in the pronounced disk feature at $\sim -130$ \kms\ should minimize contamination. 




Figure~\ref{fig:vhelio_vs_fehphot} illustrates our membership determination for H, S, D, and f130\_2 in terms of the relationship between \vhelio\ and \fehphot. We identified 73, 84, 68, and 36 RGB stars as M31 members in fields H, S, D, and f130\_2, respectively, out of 90, 89, 84, and 78 targets with successful radial velocity measurements. Using the same membership criteria as in fields H and S, we re-determined membership homogeneously for f130\_2, resulting in a final 11 star sample with reliable abundances (\S~\ref{sec:sample}) that is \textit{not} identical to the 11 star sample presented in \paperi. We included some stars that were originally excluded in \paperi\ as a consequence of lacking membership probabilities from shallow 1200G spectra (owing to failed radial velocity measurements). We excluded some stars that were originally included in \paperi\ as a result of using radial velocity as a membership diagnostic, where we did not take radial velocity into account to determine membership in \paperi\ to avoid kinematic bias.

\subsection{Kinematic Decomposition}
\label{sec:kinematic_decomp}

In Figure~\ref{fig:kinematic_decomp}, we present the heliocentric radial velocity distributions for M31 RGB stars in all four fields. We also show the full velocity distributions for stars with successful radial velocity measurements, including MW contaminants, for a total of 105, 111, 124, and 64 stars in fields H, S, D, and f130\_2. Field f130\_2 was shown to have no detected substructure by \citet{Gilbert2007}, which is consistent with our velocity distribution (see also \paperi). For fields H and S, velocity distributions have previously been analyzed in fields that contain partial overlap (Figure~\ref{fig:fields}). The mean velocity of substructure along GSS fields \citep{Gilbert2009b} and the velocity dispersion of substructure near the 12 kpc inner halo field \citep{Gilbert2007} are known to vary with radius. Thus, to compare abundances of different kinematic components within H, S, and D, it is necessary to characterize the velocity distributions of the current sample. In particular, fields S and D show clear evidence of substructure from inspection of Figure~\ref{fig:kinematic_decomp}, such as the GSS ($\sim$ $-$500 \kms) and the kinematically cold component of unknown origin \citep{Kalirai2006a, Gilbert2009b} located at approximately $-400$ \kms\ in field S, and M31's outer northeastern disk ($-130$ \kms) in field D. Although less clear, the velocity distribution of H is more strongly peaked at the systemic velocity of M31, $v_{\rm{M31}} = -300$ \kms, than expectations for a pure stellar halo component, which suggests the presence of substructure (\S~\ref{sec:se_shelf}).

We separated fields with indications of substructure--H, S, and D--into kinematically cold components and the kinematically hot stellar halo by describing the velocitiy distributions as a Gaussian mixture, such that the log likelihood function is given by,

\begin{equation}
\label{eq:lkhd}
\ln \mathcal{L} = \sum_{i=1}^{n} \ln \left( \sum_{k=1}^{K} f_k \mathcal{N} (v_i | \mu_k, \sigma_k^2) \right),
\end{equation}
where $i$ is the index representing a M31 RGB star, $v_i$ is its heliocentric radial velocity, and $n$ is the total number of M31 RGB stars in a field. $K$ is the \textit{total} number of components in a field, including the stellar halo component, where $k$ represents the index for a given component, and $f_k$ represents the normalized fractional contribution of each component to the total distribution. Each component is described by a mean velocity, $\mu_k$, and velocity dispersion, $\sigma_k$.

\begin{figure*}
\includegraphics[width=\textwidth]{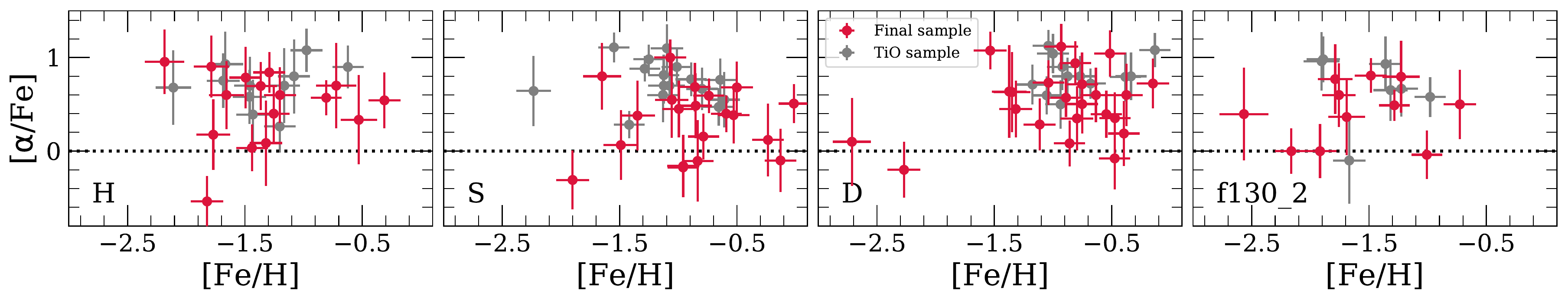}
\caption{[$\alpha$/Fe] versus [Fe/H] for RGB stars in M31 (\S~\ref{sec:abund_full}). We show abundance distributions for the inner stellar halo at 12 kpc (H), the GSS at 22 kpc (S), the outer disk at 26 kpc (D), and the smooth inner stellar halo at 23 kpc (f130\_2). We present measurements for 70 stars comprising our final sample (\textit{red circles}; \S~\ref{sec:sample}), including 46 additional M31 RGB stars with spectroscopic evidence of TiO absorption that otherwise pass our selection criteria (\textit{grey circles}). We find that all four fields are $\alpha$-enhanced ($\langle$\alphafe$\rangle$ $\gtrsim$ 0.35), and that the outer disk and GSS fields are more metal-rich on average than the 12 and 23 kpc halo fields.} 
\label{fig:m31_abund}
\end{figure*}

Given our usage of radial velocity as a diagnostic for membership (\S~\ref{sec:membership}), which excludes stars with MW-like velocities as nonmembers, the velocity distributions for M31 members in our fields are kinematically biased toward negative heliocentric velocities. As a consequence, the positive velocity tail of the stellar halo distribution in each field is truncated, such that we could not reliably fit for a halo component in each field (i.e., the velocity dispersion of the fitted halo component would likely be smaller than the true velocity dispersion of M31's stellar halo in a given region).
Therefore, we fixed the stellar halo component in each field. \citet{Gilbert2018} measured global properties of the M31 stellar halo's velocity distribution as a function of radius using over 5000 M31 RGB stars across 50 fields.
They used the likelihood of M31 membership (\S~\ref{sec:membership}; without the use of radial velocity as a diagnostic) as a prior, simultaneously fitting for all M31 and MW components. This resulted in a kinematically unbiased estimation of parameters characterizing the M31 halo's stellar velocity distribution. We transformed their mean velocities and velocity dispersions in the appropriate radial bins from the Galactocentric to heliocentric frame, based on the median right ascension and declination of all stars in a given field. Table~\ref{tab:kinematic_decomp} contains the parameters describing the heliocentric velocity distribution of the stellar halo component in each field.

We determined the number of components in each field by using an expectation-maximization (EM) algorithm to fit models of Gaussian mixtures to the velocity distribution of M31 RGB stars. Varying the number of components per model of each field, we utilized the Akaike information criterion (AIC) to select the best-fit Gaussian mixture, penalizing mixtures that did not significantly reduce the AIC without also decreasing the Bayesian information criterion (BIC). Based on this analysis, the number of components in S and D are 3 and 2, respectively, where one component in each field corresponds to the kinematically hot halo. The EM algorithm strongly preferred a single-component model for H based on the AIC and BIC. However, the velocity dispersion of this single-component model, 82 \kms, is discrepant with the velocity dispersion of M31's stellar halo between 8$-$14 kpc, 108 \kms, as measured from 525 M31 RGB stars \citep{Gilbert2018}. A two-sided Kolmogorov-Smirnov (KS) test similarly indicates that the velocity distribution of M31 RGB stars in field H is inconsistent with being solely drawn from the 8$-$14 kpc stellar halo model of \citet{Gilbert2018} at the 2\% level. Thus, we assumed a two-component model, as opposed to a single-component model, for this field. This second component likely corresponds to the inner halo substructure known as the Southeast shelf (\S~\ref{sec:se_shelf}; \citealt{Fardal2007,Gilbert2007}), where the Southeast shelf has been identified in all shallow spectroscopic fields neighboring field H (Figure~\ref{fig:fields}).

We sampled from the posterior distribution of the velocity model (Eq~\ref{eq:lkhd}) for each field using an affine-invariant Markov chain Monte Carlo (MCMC) ensemble sampler \citep{Foreman-Mackey2013}. We enforced normal prior probability distributions for $\mu_k$ and $\sigma_k$ parameters in fields H and S based on literature measurements \citep{Gilbert2018} for nearby fields (Figure~\ref{fig:fields}). 
For H, we assumed $\mu_{1,0}$ = $-$300 $\pm$ 20 \kms\ and  $\sigma_{1,0}$ = 55 $\pm$ 20 \kms, whereas for S, we assumed  $\mu_{1,0}$ = $-$490 $\pm$ 10 \kms, $\sigma_{2,0}$ = 25 $\pm$ 10 \kms, $\mu_{2,0}$ = $-$390 $\pm$ 10 \kms, and $\sigma_{2,0}$ = 20 $\pm$ 10 \kms. For field D, we assumed a flat prior, given the absence of previous modeling in the literature for the overlapping 1200G field H13d (Figure~\ref{fig:fields}). In each case, we assumed a minimum value for all dispersion parameters, $\sigma_k$, of 10 \kms, based on our typical velocity uncertainty (\S~\ref{sec:rv}). For the remainder of the bounds on each parameter, we adopted reasonable ranges that allowed for relatively unrestricted exploration of parameter space. This is intended to account for differences in the properties of our fields as compared to those of nearby fields in the literature. Additionally, we allowed $f_k$ parameters to extend down to zero for kinematically cold components.

\begin{figure*}
    \centering
    \includegraphics[width=0.85\textwidth]{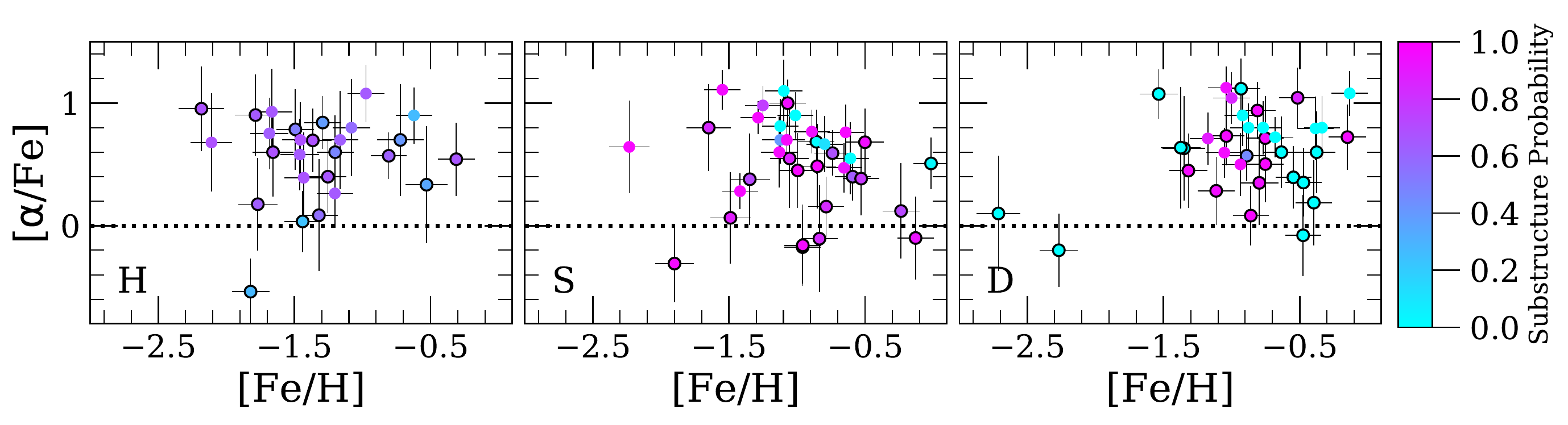}
    \caption{\alphafe\ versus \feh\ for M31 RGB stars in fields with substructure (i.e., excluding the 23 kpc smooth halo field, f130\_2), color-coded as in Figure~\ref{fig:vhelio_vs_fehphot} (\S~\ref{sec:abund_substruct}). We show M31 RGB stars in the final sample (\textit{black outlined circles}) and the TiO sample (\S~\ref{sec:sample}). 
    The abundances in the final sample of the 22 kpc GSS field (S) probe substructure almost exclusively, whereas in the 12 kpc halo field (H) and 26 kpc disk field (D), the final sample of abundances represent a mixture of the stellar halo and substructure.}
    \label{fig:m31_abund_prob}
\end{figure*}

We used 100 chains and 10$^{4}$ steps per field, for a total of 10$^{6}$ samples of the posterior probability distribution. We calculated the mean parameter values describing the velocity distribution model using the 50$^{\rm{th}}$ percentile values of the corresponding marginalized posterior probability distributions. We constructed the marginal distributions using only the latter 50\% of the MCMC chains, which are securely converged for every slitmask and model parameter in terms of stabilization of the autocorrelation time. The errors on each parameter are calculated based on the 16$^{\rm{th}}$ and 84$^{\rm{th}}$ percentiles of the marginal distributions.

\subsection{Probability of Substructure}
\label{sec:prob_substruct}

To extract the properties of the various components in each field, we assign a probability of belonging to substructure to every M31 RGB star. We 
computed the substructure probability under the 5$\times$10$^{5}$ models from the converged portion of the MCMC chain. The total probability of belonging to substructure is,

\begin{equation}
p(v_i) = \frac{ e^{\langle L_i \rangle} }{ 1 + e^{\langle L_i \rangle}},
\end{equation}
given a measurement of a star's velocity, $v_i$. $\langle L_i \rangle$ is the relative log likelihood that a M31 RGB star belongs to substructure as opposed to the stellar halo, which we express as,

\begin{equation}
\langle L_i \rangle = \ln \left( \frac{ \sum_{k=1}^{K-1} f_k \mathcal{N} ( v_i | \mu_k, \sigma_k^2 ) }{ f_{\rm{halo}} \mathcal{N} ( v_i | \mu_{\rm{halo}}, \sigma_{\rm{halo}}^2 ) } \right).
\end{equation}

Thus, we constructed a distribution function for the substructure probability in each field based on its full velocity model. For each M31 RGB star, we adopted the 50$^{\rm{th}}$ percentile value of the probability distribution function to represent the probability of the star belonging to a particular component. 

Figure~\ref{fig:vhelio_vs_fehphot} demonstrates the properties of stars likely belonging to \textit{any} substructure component in a given field in terms of heliocentric velocity and photometric metallicity. The majority of M31 RGB stars in field D belong to M31's stellar halo as opposed to its disk, whereas field S is dominated by the GSS and the kinematically cold component. In contrast, the stars in H are approximately evenly distributed between the stellar halo and substructure. If an M31 RGB star has a probability of belonging to a particular component that exceeds 50\%, i.e., it is more likely to belong to a given component than not, we associated it with the component in the subsequent abundance analysis (\S~\ref{sec:abund_substruct}).

\subsection{Resulting Velocity Distributions}
\label{sec:kinematics_results}

\begin{figure*}
    \centering
    \includegraphics[width=0.8\textwidth]{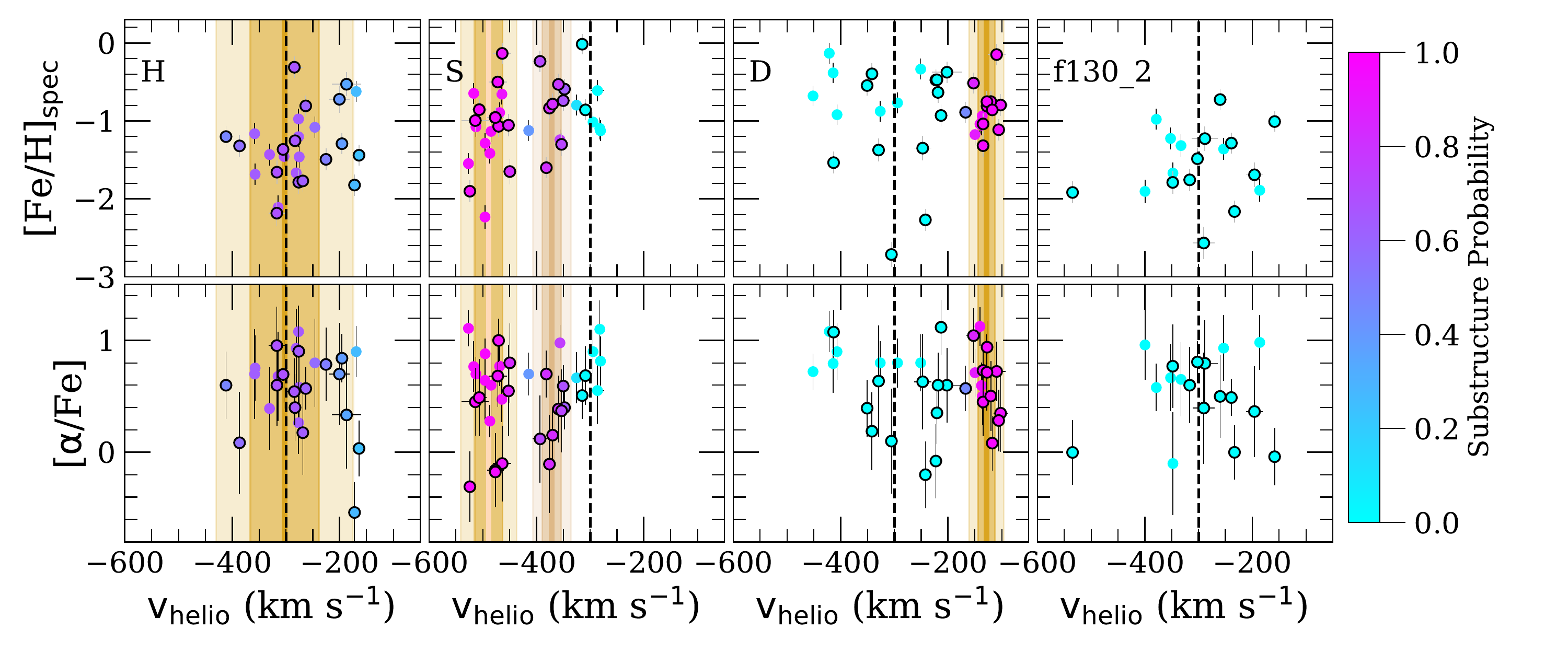}
    \caption{Spectroscopic \feh\ (\textit{top panels}) and [$\alpha$/Fe] (\textit{bottom panels}) versus heliocentric radial velocity for the same samples and color-coding as Figure~\ref{fig:m31_abund_prob} (\S~\ref{sec:abund_substruct}), except including the 23 kpc smooth halo field (f130\_2). The dashed vertical lines and bands are the same as Figure~\ref{fig:vhelio_vs_fehphot}, representing the median parameters of the velocity distributions in each field (\S~\ref{sec:kinematics_results}). The substructure in the 12 kpc halo field (H) is difficult to distinguish from the stellar halo in terms of \feh\ and \alphafe, where the same is true between the GSS core and KCC of unknown origin in the 22 kpc GSS field (S). M31's disk in the 26 kpc disk field (D) appears narrow in \feh\ relative to the stellar halo.}
    \label{fig:vhelio_vs_fehspec_alphafe}
\end{figure*}

We summarize the mean velocity distribution model parameters for fields H, S, D, and f130\_2 in Table~\ref{tab:kinematic_decomp} and illustrate the multiple-component models for each field in Figure~\ref{fig:kinematic_decomp}. For H, we identified a relatively cold component with $\langle v \rangle$ = $-295$ \kms\ and $\sigma_v$ = 66 \kms, which we attribute to the Southeast shelf (\S~\ref{sec:se_shelf}; \citealt{Fardal2007,Gilbert2007}), a tidal shell originating from the GSS progenitor. The fractional contribution of this component is uncertain, ranging from 0.3$-$0.8, and exhibits covariance with the velocity dispersion, where increasing (decreasing) the fractional contribution of the substructure component increases (decreases) its velocity dispersion. Substructure components are more robustly characterized in fields S and D. We find that $\langle v \rangle$ = $-$489 \kms, $\sigma_v$ = 26 \kms\ for the GSS, additionally recovering the secondary kinematically cold component of unknown origin \citep{Kalirai2006a,Gilbert2009b, Gilbert2019} separated by $\sim$ $-$120 \kms\ in line-of-sight velocity ($\langle v \rangle$ = $-$372 \kms, $\sigma_v$ = 17 \kms) from the primary GSS feature.\footnote{Relative to previous determinations of the velocity distribution in the 22 kpc field \citep{Gilbert2018}, the KCC is offset toward lower mean heliocentric velocities by $\sim$20 \kms. This may result from the reduced velocity precision of the 600ZD grating (\S~\ref{sec:rv}), or alternatively, differences in spatial configuration of the sample.} 

For M31's northeastern disk, we find $\langle v \rangle$ = $-$128 \kms, $\sigma_v$ = 16 \kms, indicating that the disk rotation velocity is 191 \kms\ offset from M31's halo velocity in this field.\footnote{We acknowledge the possibility of bias introduced into our measurement as a result of the $-$100 \kms\ velocity cut utilized in our membership determination for the disk field (\S~\ref{sec:membership}). If we have excluded a significant fraction of M31 RGB stars redshifted to low heliocentric velocity as a consequence of the disk rotation, then our measurements for the disk would underestimate the mean velocity and velocity dispersion (Appendix~\ref{sec:disk_dispersion}).} For a comparison of the dispersion the outer disk feature with the literature, see Appendix~\ref{sec:disk_dispersion}. The peak of our disk velocity distribution, $v$ = $-$128 \kms, agrees with previous studies of disk kinematics along the northeast major axis, which measured line-of-sight velocities of $\sim-$100 \kms\ for fields along the major axis \citep{Ibata2005, Dorman2012}. However, we note that  field D ($r_{\rm{maj}}$ = 25.6 kpc) is located beyond the maximum major axis distance probed by these studies. Although M31's disk is a prominent feature, field D is dominated by the kinematically hot stellar halo component ($f_{\rm{halo}}$ = 0.57).

 Assuming a simple model \citep{Guhathakurta1988} for perfectly circular rotation of an inclined disk ($i$ = 77$^\circ$, P.A. = 38$^\circ$), the line-of-sight mean velocity of the disk feature corresponds to $v_{\rm{rot}}$ = 229$-$244 \kms\ in the disk plane. Based on a rotation curve inferred from \ion{H}{1} kinematics between 10$-$30 kpc and corrected for the inclination of M31's disk \citep{Klypin2002,Ibata2005}, the expected circular velocity at field D ($r_{\rm{disk}}$ = 35 kpc)  is $\sim$240 \kms, corresponding to a line-of-sight velocity of $-$119 \kms\ \citep{Guhathakurta1988}. Thus, we computed the expected deviation from perfectly circular rotation, $v_{\rm{lag}}$, for the disk feature in field D. Accounting for uncertainty in the mean velocity of the disk feature resulting from the fitting procedure and the membership determination,  we estimated that $v_{\rm{lag}}$ = $-$9$^{+11}_{-3}$ \kms. For RGB stars in M31's disk between $\sim$5-15 kpc, \citet{Quirk2019} found that $\langle v_{\rm{lag}} \rangle$ $\sim$ 63 \kms, although our inferred value is not inconsistent with their full $v_{\rm{lag}}$ distribution.

\section{Elemental Abundances of the M31 Fields}
\label{sec:abund}

In \S~\ref{sec:kinematics}, we modeled the velocity distributions of the 12 kpc inner halo (H), 22 kpc GSS (S), 26 kpc outer disk (D), and 23 kpc smooth halo (f130\_2) fields, identifying substructure in the first three fields. Hereafter, we refer to the 12 kpc substructure as the SE shelf (\S~\ref{sec:se_shelf}), the primary 22 kpc substructure as the GSS core, the secondary 22 kpc substructure as the KCC, and the 26 kpc substructure as the disk, for clarity of interpretation when analyzing the abundance distributions. A catalog of stellar parameters and elemental abundances for individual M31 RGB stars across the 4 fields is contained in Appendix~\ref{sec:abund_table}.

\subsection{Full Abundance Distributions}
\label{sec:abund_full}

\begin{figure}
    \centering
    \includegraphics[width=\columnwidth]{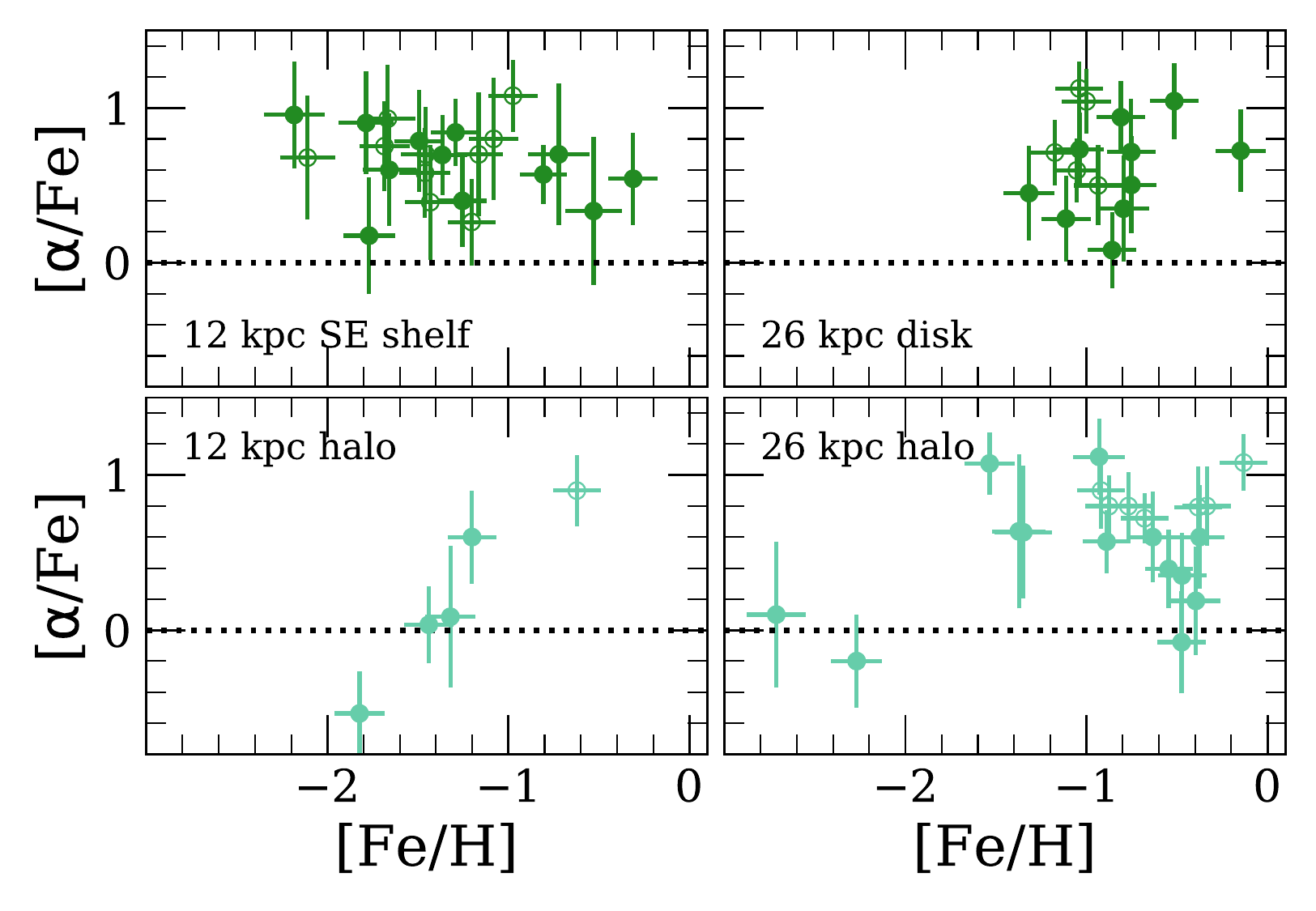}
    \caption{\alphafe\ versus \feh\ for M31 RGB stars with $\delta$([$\alpha$/Fe]) $<$ 0.5 in the 12 kpc inner halo field (H; \S~\ref{sec:abund_H})  and 26 kpc outer disk field (D; \S~\ref{sec:abund_D}). We separated each field into its kinematic components by assigning stars to the component to which it has the highest probability of belonging, based on its modeled velocity distribution (\S~\ref{sec:prob_substruct}). Stars with TiO absorption (\S~\ref{sec:sample}) are represented as open circles. We show abundances of M31 RGB stars in the SE shelf feature (\textit{upper left}), M31's disk (\textit{upper right}), and M31's stellar halo (\textit{bottom panels}).}
    \label{fig:HD_abund_decomp}
\end{figure}

We present [$\alpha$/Fe] versus [Fe/H] for 70 M31 RGB stars across the 12 kpc halo field, 22 kpc GSS field, 26 kpc outer disk field, and 23 kpc smooth halo field in Figure~\ref{fig:m31_abund}. We also show 46 M31 RGB stars with TiO absorption that otherwise pass our selection criteria (\S~\ref{sec:sample}). Table~\ref{tab:m31_avg_abund} summarizes the \feh\ and \alphafe\ abundances for all M31 RGB stars in our final sample (i.e., without TiO, $\delta$(\feh) $<$ 0.5, and $\delta$([$\alpha$/Fe]) $<$ 0.5) in each field. Given that we have a finite sample subject to bias, we performed bootstrap re-sampling (with 10$^{4}$ draws) to estimate mean abundances and abundance spreads for each field, including 68\% confidence intervals on each parameter. Since the percentage of M31 RGB stars affected by TiO absorption across all four fields is similar, we anticipate that the relative metallicity differences between fields are accurate. Figure~\ref{fig:m31_avg_all} provides a visual representation of the data in Table~\ref{tab:m31_avg_abund}, where we have included equivalent measurements of $\langle$\feh$\rangle$ and $\langle$\alphafe$\rangle$ in M31 RGB stars in the outer halo \citep{Vargas2014b} and a 17 kpc GSS field (\paperii). 

On average, we find that our M31 sample is $\alpha$-enhanced (0.40 $\lesssim$ $\langle$[$\alpha$/Fe]$\rangle$ $\lesssim$ 0.60) and spans a metallicity range of $-$1.5 $\lesssim$ $\langle$[Fe/H]$\rangle$ $\lesssim$ $-$0.9. High $\alpha$-element abundances indicate that the stellar populations in our M31 fields, regardless of the various galactic structures to which they belong, are likely characterized by rapid star formation and dominated by the yields of core-collapse supernovae. 
The range of $\langle$[Fe/H]$\rangle$ indicates a range of star formation duration. Additionally, stars in all four fields possess a similar spread in \feh ($\sim$0.47-0.55), supporting either extended star formation for a single origin, or a multiple-progenitor hypothesis. The GSS field and outer disk fields are the most metal-rich, suggesting \added{either} more extended \added{or efficient} SFHs compared to the 12 kpc and 23 kpc stellar halo fields. Considering simple field averages, stars in the GSS field and outer disk field are indistinguishable from one another in terms of [Fe/H]. Interestingly, the GSS field may be less $\alpha$-enhanced than the 26 kpc disk field, with a difference in $\langle$\alphafe$\rangle$ of 0.17$^{+0.11}_{-0.12}$. If so, this suggests different relative star formation timescales between Types Ia and \replaced{II}{core-collapse} supernovae, or differences in star formation efficiency, between M31's outer disk and the GSS progenitor. In accordance with expectations of stellar halo formation, the 23 kpc smooth halo field appears to be more metal-poor than the 12 kpc halo field, by 0.24 $\pm$ 0.18 dex on average. We discuss the possibility of \added{radial} abundance gradients, in both \feh\ and \alphafe, in the stellar halo of M31 in \S~\ref{sec:abund_gradient}.

\subsection{Abundance Distributions of Individual Kinematic Components}
\label{sec:abund_substruct}

\begin{figure}
    \centering
    \includegraphics[width=2in, height=3in, keepaspectratio]{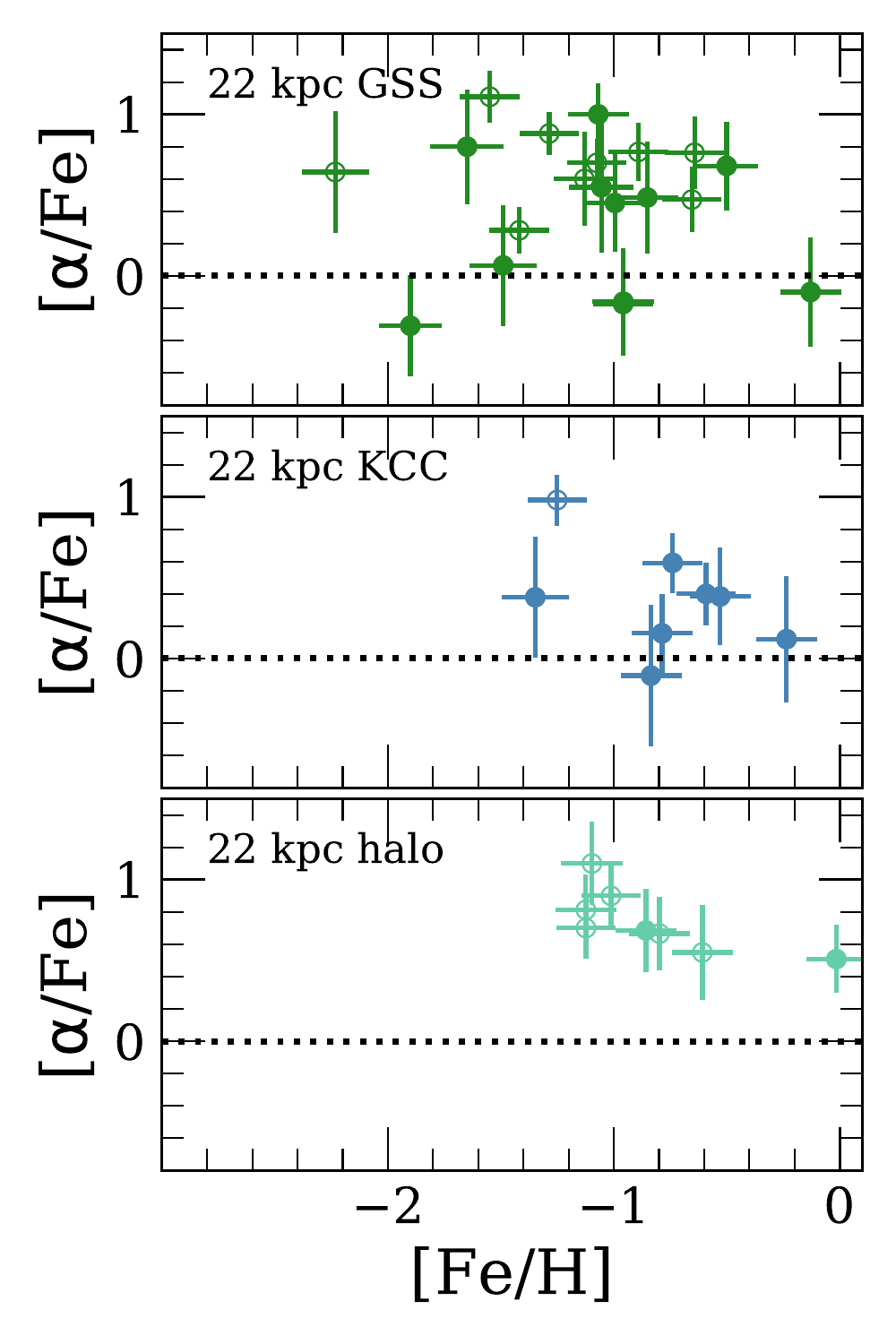}
    \caption{Same as Figure~\ref{fig:HD_abund_decomp}, except for the 22 kpc GSS field (S; \S~\ref{sec:abund_S}). We show abundances of M31 RGB stars in the GSS (\textit{top panel}), the KCC of unknown origin (\textit{middle panel}), and M31's stellar halo (\textit{bottom panel}).}
    \label{fig:S_abund_decomp}
\end{figure}

Given that we have identified substructure (\S~\ref{sec:kinematics_results}) in the 12 kpc halo field, 22 kpc GSS field, and 26 kpc disk field, we separate the full abundance distributions (\S~\ref{sec:abund_full}) into the underlying kinematic components. Using the modeled velocity distributions, we assign each M31 RGB star in fields with substructure a probability of belonging to each individual component (\S~\ref{sec:prob_substruct}). Figure~\ref{fig:m31_abund_prob} shows \alphafe\ versus \feh\ for the 12 kpc halo, 22 kpc GSS, and 26 kpc disk fields, where we have indicated the probability that an individual M31 RGB star belongs to \textit{any} substructure component. Our abundance measurements in the 22 kpc GSS field probe substructure almost exclusively, whereas the abundances in the 12 kpc  halo and 26 kpc disk fields represent a mixture of the stellar halo and substructure. Figure~\ref{fig:vhelio_vs_fehspec_alphafe} shows the probabilistic distributions of \feh\ and \alphafe\ for each kinematic component, where we have plotted \alphafe\ and \feh\ against heliocentric velocity. At a glance, the SE shelf is difficult to chemically distinguish from the stellar halo, where this statement also applies between the GSS core and KCC. M31's disk appears narrow in \feh\ relative to the stellar halo.

\addtolength{\tabcolsep}{-2.5pt}
\begin{table}
    \centering
    \begin{threeparttable}
    \caption{Abundances in M31 Fields}
    \begin{tabular*}{\columnwidth}{lcccc}
        \hline
        \hline
        Comp.\tnote{a} & $\langle$\feh$\rangle$\tnote{b} & $\sigma$(\feh)\tnote{c} & $\langle$\alphafe$\rangle$ & $\sigma$(\alphafe) \\ \hline
        \multicolumn{5}{c}{12 kpc Halo Field (H)} \\\hline
        Field\tnote{d} & $-$1.30$^{+0.12}_{-0.11}$ & 0.47 $\pm$ 0.08 &0.50$^{+0.10}_{-0.11}$ & 0.38$^{+0.09}_{-0.13}$\\[0.4em]
        SE Shelf & $-$1.30$^{+0.13}_{-0.12}$ & 0.49$^{+0.08}_{-0.09}$ &0.53$^{+0.08}_{-0.10}$ & 0.36$^{+0.09}_{-0.11}$ \\[0.4em]
        Halo & $-$1.30 $\pm$ 0.11 & 0.45$^{+0.07}_{-0.08}$ &0.45$^{+0.12}_{-0.13}$ & 0.42$^{+0.09}_{-0.14}$\\[0.1em]
        \hline
        \multicolumn{5}{c}{22 kpc GSS Field (S)} \\ \hline
        Field & $-$0.84 $\pm$ 0.10 & 0.46$^{+0.07}_{-0.08}$ & 0.41$^{+0.08}_{-0.09}$ & 0.35$^{+0.06}_{-0.05}$\\[0.4em]
        GSS & $-$1.02$^{+0.15}_{-0.14}$ & 0.45$^{+0.10}_{-0.11}$ &0.38$^{+0.17}_{-0.19}$ & 0.45$^{+0.07}_{-0.08}$\\[0.4em]
        KCC & $-$0.71 $\pm$ 0.11 & 0.27 $\pm$ 0.09 & 0.35$^{+0.08}_{-0.09}$ & 0.18$^{+0.04}_{-0.05}$\\[0.4em]
        Halo & $-$0.66$^{+0.16}_{-0.18}$ & 0.44$^{+0.07}_{-0.10}$ &0.49$^{+0.05}_{-0.06}$ & 0.21$^{+0.05}_{-0.04}$\\[0.1em]
        \hline
        \multicolumn{5}{c}{26 kpc Disk Field (D)} \\ \hline
        Field & $-$0.92$^{+0.10}_{-0.12}$ & 0.55$^{+0.11}_{-0.12}$ &0.58 $\pm$ 0.08 & 0.36$^{+0.04}_{-0.05}$\\[0.4em]
        Disk & $-$0.82 $\pm$ 0.09 & 0.28$^{+0.07}_{-0.09}$ & 0.60$^{+0.09}_{-0.10}$ & 0.28$^{+0.05}_{-0.06}$\\[0.4em]
        Halo & $-$1.00$^{+0.17}_{-0.19}$ & 0.68$^{+0.12}_{-0.14}$ &0.55 $\pm$ 0.13 & 0.40$^{+0.06}_{-0.08}$\\[0.1em]
        \hline
        \multicolumn{5}{c}{23 kpc Halo Field (f130\_2)} \\ \hline
        Field & $-$1.54 $\pm$ 0.14 & 0.47$^{+0.08}_{-0.09}$ &0.43$^{+0.11}_{-0.12}$ & 0.31 $\pm$ 0.05 \\[0.1em]
        \hline \hline
    \end{tabular*}
    \begin{tablenotes}
    \item Note.\textemdash\ All quantities are calculated from bootstrap resampling of the final sample. For a discussion of bias in the sample, see \S~\ref{sec:sample_select}. (a) For the components of each field, measurements are additionally weighted by the probability of belonging to a given component (\S~\ref{sec:prob_substruct},~\ref{sec:abund_substruct}) (b) Inverse-variance weighted mean. (c) Inverse-variance weighted standard deviation. (d) ``Field'' refers to all M31 RGB stars present in a field, regardless of association with a kinematic component. 
    \end{tablenotes}
    \label{tab:m31_avg_abund}
    \end{threeparttable}
\end{table}
\addtolength{\tabcolsep}{2.5pt}

Figures~\ref{fig:m31_abund_prob} and ~\ref{fig:vhelio_vs_fehspec_alphafe} emphasize that the association of a M31 RGB star with any given component is not definitive. Thus, when computing $\langle$\feh$\rangle$ and $\langle$\alphafe$\rangle$ for each component (Table~\ref{tab:m31_avg_abund}), we weighted each abundance measurement by its probability of belonging to a particular component, in addition to weighting by the inverse variance of the measurement uncertainty. For clarity of illustration, Figures~\ref{fig:HD_abund_decomp} and~\ref{fig:S_abund_decomp} show \feh\ and \alphafe\ abundances for the kinematic components in each of the three fields with substructure, where we have assigned each star to the component to which it is most likely to belong (\S~\ref{sec:prob_substruct}). The M31 RGB stars in the final abundance sample of the 12 kpc and 26 kpc fields represent the relative fraction of the stellar halo and substructure components (Table~\ref{tab:kinematic_decomp}) accurately. In contrast, M31 RGB stars in the final abundance sample the 22 kpc field under-represent the estimated stellar halo fraction by $\sim$10\% and over-represent the KCC.

\begin{figure}
    \centering
    \includegraphics[width=\columnwidth]{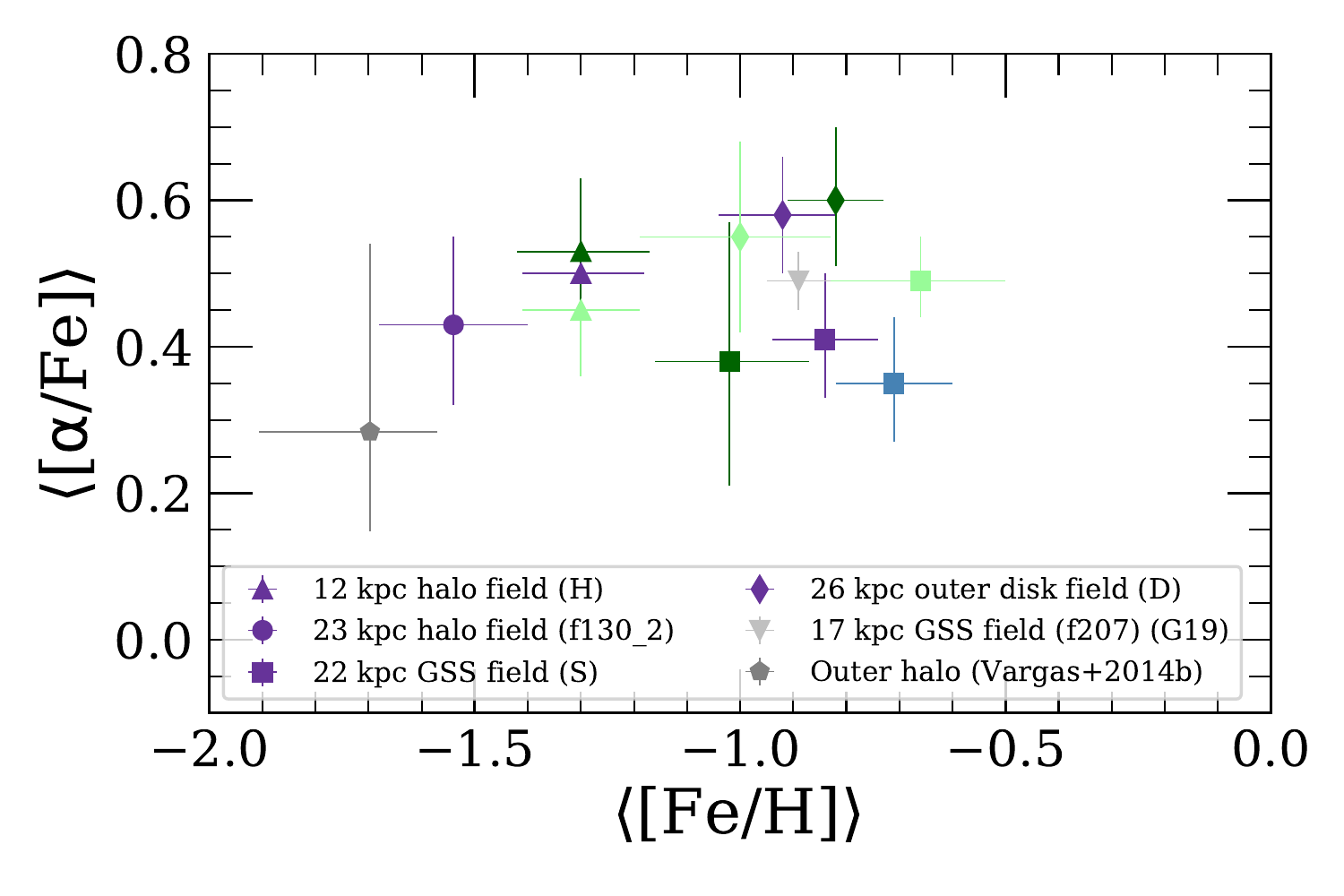}
    \caption{$\langle$\alphafe$\rangle$ versus $\langle$\feh$
    \rangle$ for all M31 fields (\S~\ref{sec:abund}). The data are presented in Table~\ref{tab:m31_avg_abund}. We show the averages for the entire field (\textit{purple}; \S~\ref{sec:abund_full}), regardless of kinematic component, in addition to the probabilistic average for each kinematic component (\S~\ref{sec:abund_substruct}): the stellar halo (\textit{light green}), the primary KCC (\textit{dark green}), and the secondary KCC for the GSS field (\textit{blue}). We overplot average abundance measurements from similarly deep spectra of M31 RGB stars ({\it grey points}) in a 17 kpc GSS field (\paperii) and four outer halo stars between $\sim$70-140 kpc \citep{Vargas2014b}.}
    \label{fig:m31_avg_all}
\end{figure}

In addition to representing field averages, Figure~\ref{fig:m31_avg_all} shows the average probabilistic \feh\ and \alphafe\ for each kinematic component in the three M31 fields with substructure. 
The bias against red stars, which are presumably more metal-rich, largely incurred by the omission of TiO stars (\S~\ref{sec:sample_select}) affects the final abundance distribution of the SE shelf and GSS core disproportionately relative to other kinematic components present in the 12 kpc and 22 kpc fields (Figure~\ref{fig:m31_cmd}). We also note that there is a population of stars falling on the solar metallicity isochrone attributed to the KCC for which we were unable to measure abundances. We anticipate that the difference in $\langle$[Fe/H]$\rangle$ between the SE shelf and 12 kpc stellar halo may be larger than the quoted values (Table~\ref{tab:m31_avg_abund}), whereas it is difficult to predict how these effects would impact the abundances of the GSS core compared to the KCC. An equivalent number of M31 RGB stars in both the disk and 26 kpc stellar halo were omitted from the final sample, such that the chemical composition of each component should be similarly impacted.

\subsubsection{12 kpc Halo Field}
\label{sec:abund_H}

For the 12 kpc halo field, we find that $\langle$[Fe/H]$\rangle$ and $\langle$\alphafe$\rangle$ for the SE shelf cannot be statistically distinguished from the stellar halo (Table~\ref{tab:m31_avg_abund}). Although we weighted our field sample by substructure probability computed from the velocity distribution, stars that are more likely to belong to the SE shelf ($p > 0.5$; \S~\ref{sec:prob_substruct}) still have an average probability of 35\% of belonging to the stellar halo. Considering that our final sample for this field does not include many of the red stars that are more likely to populate the SE shelf (Figure~\ref{fig:m31_cmd}), it is possible that the SE shelf is more metal-rich than the halo. Given the uncertainty on $\langle$[$\alpha$/Fe]$\rangle$,  the SE shelf and stellar halo may be similarly $\alpha$-enhanced, or the SE shelf may in fact be more $\alpha$-rich than the halo. 
We discuss the possibility that the SE shelf is related to the GSS progenitor in \S~\ref{sec:se_shelf}.


\subsubsection{22 kpc GSS Field}
\label{sec:abund_S}

\begin{figure}
    \centering
    \includegraphics[width=0.85\columnwidth]{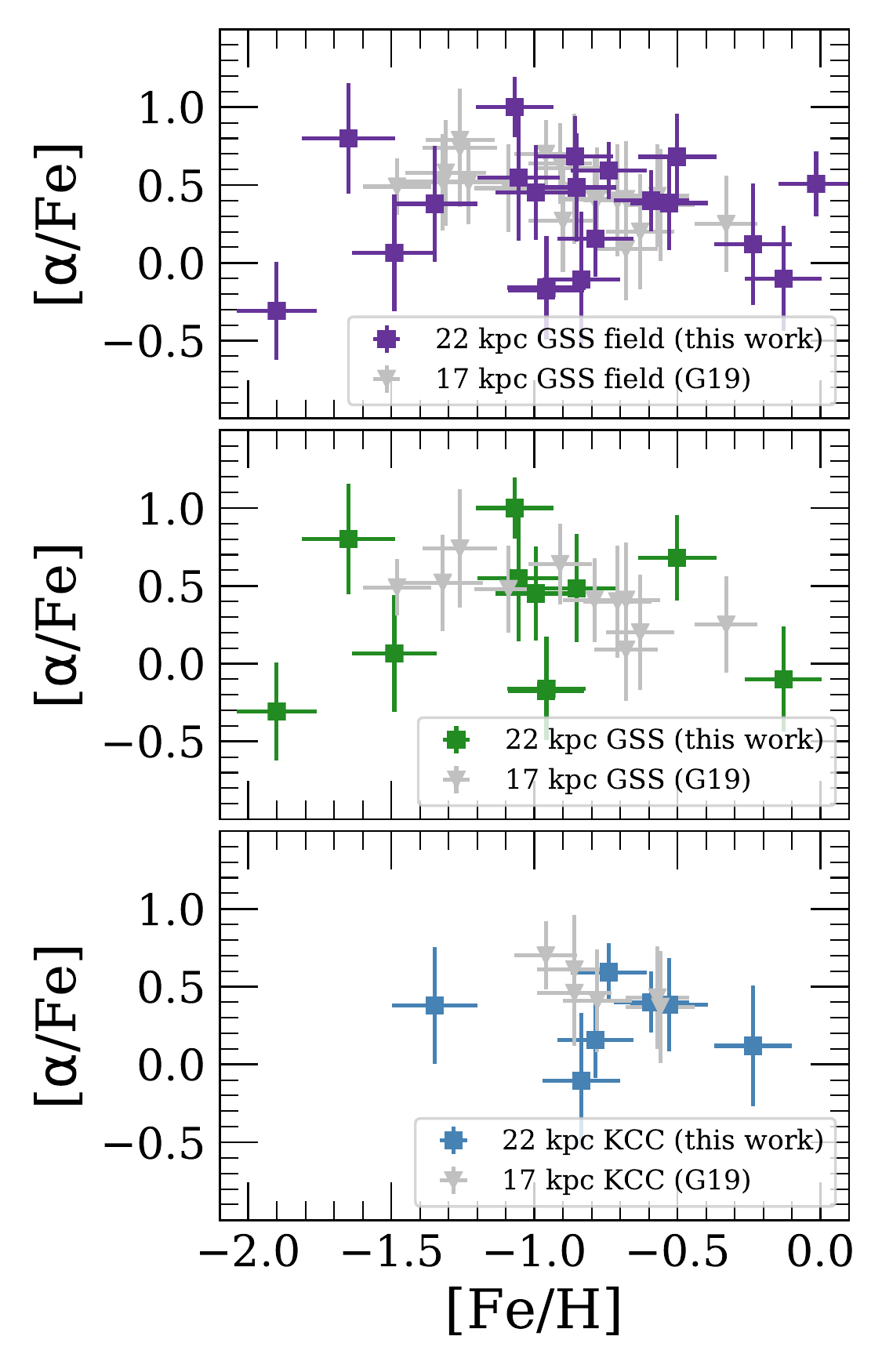}
    \caption{\alphafe\ versus \feh\ for M31 RGB stars in the 22 kpc GSS field (S; {\it colored squares}, \S~\ref{sec:abund_S}) compared to a 17 kpc GSS field ({\it grey triangles}; \paperii). We present abundances for all M31 RGB stars in a given field ({\it top panel}), the GSS core ({\it middle panel}), and the KCC of unknown origin ({\it bottom panel}). We find that the abundance distributions for the GSS between 17 and 22 kpc are consistent.}
    \label{fig:gss_S_f207}
\end{figure}

When separating the GSS core from the KCC, we do not find evidence of a decline of \alphafe\ with \feh\ for the GSS or the KCC. Many of the RGB stars populating the apparent gradually declining \alphafe\ plateau of the 22 kpc GSS field when considered as a whole (Figure~\ref{fig:m31_abund_prob}) have a higher probability of belonging to the KCC.
We cannot identify the characteristic ``knee'' in the \alphafe\ vs. \feh\ distribution based on our abundances for the GSS core. However, the 22 kpc GSS core abundance distribution overlaps with that of a 17 kpc GSS field (Figure~\ref{fig:gss_S_f207}), where the ``knee'' is located at \feh\ $\sim-$0.9 (\paperii). Taking into account observational uncertainty, computing the intrinsic dispersion (not to be confused with the standard deviation) of the \feh\ and \alphafe\ distributions yields 0.46$^{+0.24}_{-0.13}$ and $\leq$ 0.46, respectively, for the 22 kpc GSS field and 0.28$^{+0.15}_{-0.08}$ and 0, respectively, for the 17 kpc GSS field. Based on this, we can conclude that the intrinsic dispersion of the abundance distributions between the 22 kpc and 17 kpc GSS fields are marginally consistent. Thus, the GSS abundance distributions do not differ substantially in \feh\ and \alphafe\ across the $\sim$16$-$23 kpc radial range probed by the two fields along the GSS core.

We find that the GSS core in the 22 kpc GSS field may be more metal-poor than the KCC by 0.31$^{+0.19}_{-0.18}$ dex on average, with the caveat of bias against red stars in the GSS core. For the 17 kpc GSS field, \paperii\ found that the KCC differed in $\langle$\feh$\rangle$ from the GSS core by 0.14$^{+0.54}_{-0.59}$ based on probabilistic \feh \ distributions computed from their velocity model.
Using a two-sample KS test, we found that the \alphafe\ distributions of the GSS core ($p_{\rm GSS} >$ 0.5; \S~\ref{sec:prob_substruct}) and KCC ($p_{\rm KCC} >$ 0.5) are statistically consistent in the 22 kpc GSS field, whereas the \feh\ distributions are inconsistent at the 2\% level.

The stellar halo in the 22 kpc GSS field appears to be more metal-rich than the GSS core and KCC, although the uncertainty in $\langle$[Fe/H]$\rangle$ is large. This is because our final sample in the 22 kpc GSS field over-represents substructure and provides poor constraints on the stellar halo in this region (Figure~\ref{fig:vhelio_vs_fehspec_alphafe}). \paperii\ similarly found that they could not constrain the \alphafe\ vs. \feh\ abundance distribution of the stellar halo in the vicinity of the GSS at 17 kpc, owing to insufficient sample size. However, \feh\ for the 22 kpc stellar halo is consistent with \paperii's probabilistic MDF for the 17 kpc stellar halo along the GSS. 

\subsubsection{26 kpc Disk Field}
\label{sec:abund_D}

When separating the 26 kpc disk field into the stellar halo and outer disk, we found that the disk and halo are similar in $\langle$[$\alpha$/Fe]$\rangle$ and $\langle$[Fe/H]$\rangle$, where the disk is slightly more metal-rich. However, much of this difference is driven by the two halo stars at low \feh\ ($\lesssim-$2). Omitting these two stars, we found that $\langle$[Fe/H]$\rangle_{\rm{halo}}$ = $-$0.78$^{+0.11}_{-0.13}$ and $\langle$[$\alpha$/Fe]$\rangle_{\rm{halo}}$ = 0.63$^{+0.12}_{-0.13}$. The metal-rich nature of the disk relative to the halo is not preserved in this case. It is unclear if the metal-poor stars are outliers or representative of a metal-poor tail of the halo distribution that was not well-sampled by our target selection. Given their M31-like velocities ($v_{\rm{helio}}$ $<$ $-$200 \kms; Figure~\ref{fig:vhelio_vs_fehspec_alphafe}), it is unlikely that these stars are MW foreground dwarf stars. We compare our abundances to the literature for the disks of M31 and MW in \S~\ref{sec:disk_abund}.


\section{Discussion}
\label{sec:discuss}

\subsection{Chemical Differences Between the Inner and Outer Halo of M31 and the MW}
\label{sec:abund_gradient}

We investigated whether the \feh\ and \alphafe\ abundances in our four M31 fields, combined with data from the outer halo of M31 (\citealt{Vargas2014b}), provides evidence for \added{radial} chemical abundance gradients in the stellar halo of M31. Previous studies have established the existence of a global \added{radial} metallicity gradient in M31's stellar halo based on spectroscopic \citep{Kalirai2006b, Koch2008, Gilbert2014} and photometric \citep{Ibata2014} samples of individual stars, although metallicity measurements have been primarily CMD-based with small samples of calcium-triplet based measurements. In particular, \citet{Gilbert2014} used the largest spectroscopically confirmed data set to date to analyze the CMD-based metallicity distribution of the stellar halo, with over 1500 M31 halo stars across 38 fields and detections extending beyond 100 kpc. 

\begin{figure}
    \centering
    \includegraphics[width=\columnwidth]{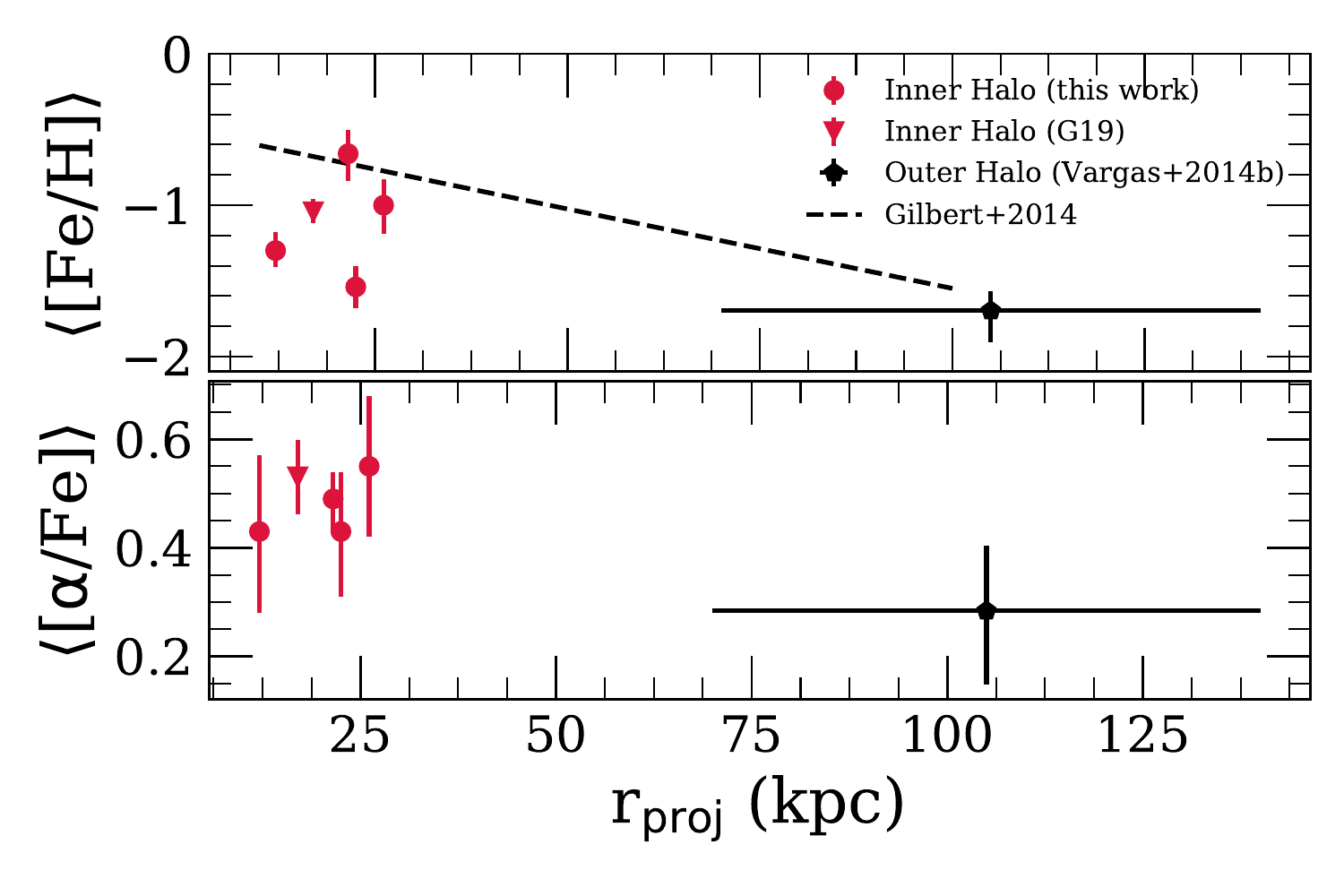}
    \caption{$\langle$\feh$\rangle$ ({\it top}) and $\langle$\alphafe$\rangle$ ({\it bottom}) as a function of projected radius in the stellar halo (i.e., with substructure removed) of M31 (\S~\ref{sec:abund_gradient}). We show the four fields presented in this work (\textit{red circles}), a 17 kpc GSS field ({\it red triangle}; \paperii), and an average for the four outer halo stars of \citet{Vargas2014b} (70$-$140 kpc) placed at 105 kpc. The inner halo fields are biased against red stars with more metal-rich \fehphot\ (\S~\ref{sec:sample_select}).
    The dashed black line represents the \deleted{stellar halo} photometric \added{radial} metallicity gradient of \added{M31's stellar halo of} \citet{Gilbert2014}, $-$0.011 dex kpc$^{-1}$, where substructure has been removed, assuming a normalization of $\langle$\feh$\rangle_{\rm{phot}}$ = $-$0.5.
    If the \citet{Vargas2014b} halo stars are representative of the outer halo, we find tentative evidence of a negative \added{radial} \alphafe\ gradient between the inner and outer halo of M31.}
    \label{fig:rproj_vs_abund}
\end{figure}

Figure~\ref{fig:rproj_vs_abund} illustrates $\langle$[Fe/H]$\rangle$ and $\langle$[$\alpha$/Fe]$\rangle$ as a function of projected radius from the center of M31 for the stellar halo component (\S~\ref{sec:kinematic_decomp}) in each field. We referred to the stellar halo components in each field as belonging to the ``inner halo'' based on their projected radius ($r_{\rm{proj}}$ $<$ 30 kpc), as opposed to any definition based on structural properties of the halo \citep{Dorman2013}. We probabilistically removed substructure from each field in order to probe the properties of the ``smooth'' stellar halo. For comparison, we show the stellar halo (i.e., with substructure removed) \added{radial} metallicity gradient of \citet{Gilbert2014}, $-$0.011 $\pm$ 0.0013 dex kpc$^{-1}$, assuming a normalization of $\langle$\feh$\rangle_{\rm{phot}}$ = $-$0.5. Owing to the exclusion of red stars with signatures of TiO in their spectra from our final sample, the inner halo fields (including the 17 kpc GSS field; \paperii) are biased toward lower \fehphot\ (\S~\ref{sec:sample_select}).  Figure~\ref{fig:rproj_vs_abund} also includes data for the 4 M31 outer halo stars of \citet{Vargas2014b}, which span a large radial range (70$-$140 kpc), shown at $r_{\rm{proj}}$ = 105 kpc. Measurements of \feh\ and \alphafe\ from spectral synthesis appear to support the existence of negative \added{radial} abundance gradients in M31's stellar halo, although larger samples of data in the outer halo are necessary to confirm this possibility.


Theoretical studies of stellar halo formation \citep{Font2011, Tissera2014, DSouzaBell2018a, Monachesi2019} have shown that M31's negative \added{radial} metallicity gradient is relatively steep compared to predictions from typical simulations. Based on such comparisons, \citet{Gilbert2014} speculated that the magnitude of M31's \added{radial} metallicity gradient implies that, in addition to a population of stars \added{having} formed \textit{in situ} in the inner regions, massive progenitors have contributed significantly to the formation of the halo. Additionally, spatial and chemical field-to-field variation in the outer halo \citep{Gilbert2012, Gilbert2014} suggests that less massive progenitors are the dominant contributors in this region.

Comparatively few theoretical studies have explored the relationship between \added{radial} gradients in \alphafe\ and accretion history in detail. \citet{Font2006} found no large-scale \added{radial} \feh\ or \alphafe\ gradients in their hierarchically formed stellar halos, which they attributed to their simulated stellar halos being dominated by early accretion in both the inner and outer halo. Including contributions from stellar populations formed \textit{in situ}, \citet{Font2011} found ubiquitously negative \added{radial} \feh\ gradients and largely flat \added{radial} \alphafe\ gradients in their simulated stellar halos. They ascribed the lack of a \alphafe\ trend to the prevalence of \replaced{Type II}{core-collapse} supernovae at all radii for both \textit{in situ} and accreted stellar halo components, which is a consequence of the typically old stellar age ($\sim$11-12 Gyr) of the latter component. A globally $\alpha$-poor outer halo would likely be caused by progenitors accreted at late epochs \citep{Robertson2005,Font2006,Johnston2008}. Thus, if the stellar halo of M31 possesses both negative \added{radial} \feh\ and \alphafe\ gradients, it may be a consequence of the contrast between massive, $\alpha$-enhanced progenitors and/or \textit{in situ} star formation dominating the inner halo and less massive, chemically evolved progenitors dominating the outer halo.

Similar to M31, the MW exhibits indications of negative \added{radial} metallicity and $\alpha$-element abundance gradients \added{(c.f. \citet{Conroy2019} concerning the MW halo's radial \feh\ gradient)}. The peaks of the MDFs of the MW's inner and outer halo correspond to \feh\ $\sim$ $-$1.5 and \feh\ $\sim$ $-$2, respectively \citep{Carollo2007,Carollo2010, deJong2010, An2013, Fernandez-Alvar2017}.  Stellar populations with distinct $\alpha$-element abundances have been identified for stars with halo-like kinematics  \citep{Fulbright2002,Gratton2003,Roederer2009, Ishigaki2010,NissenSchuster2010,Ishigaki2012, Ishigaki2013, Hawkins2015,Hayes2018b}. As opposed to relying on a kinematic decomposition, \citet{Fernandez-Alvar2015, Fernandez-Alvar2017} examined the variation of \feh\ and \alphafe\ as a function of galactocentric radius, confirming that the low-$\alpha$ population 
is associated with the outer halo ($r_{\rm{GC}}$ $>$ 15 kpc) of the MW. The dichotomy in \alphafe\ and \feh, respectively, between the inner and outer halo in the MW has generally been interpreted to mean that its outer halo corresponds to an accreted population with extended SFHs, whereas its inner halo was constructed by stars formed \textit{in situ} and/or stars accreted from chemically distinct progenitor(s).

In comparison to the MW, the metallicity of individual RGB stars attributed to the metal-poor component of M31's inner stellar halo (\feh\ $\sim$ $-$1.5; \paperi) and the outer halo of M31 (\feh\ $\sim$ $-$1.7; \citealt{Vargas2014b}) suggest that both the ``smooth'' inner halo and the outer halo of M31 are more metal-rich on average at a given projected radius than the MW. The stellar halo of M31 also appears to be $\alpha$-enhanced at all radii compared to the MW, only approaching MW halo-like \alphafe\ at large radii in M31. 

\subsection{Constructing the Inner Stellar Halo of M31 from Present-Day M31 Satellite Galaxies}
\label{sec:halo_from_dwarfs}

\begin{figure*}
    \centering
    \includegraphics[width=\textwidth]{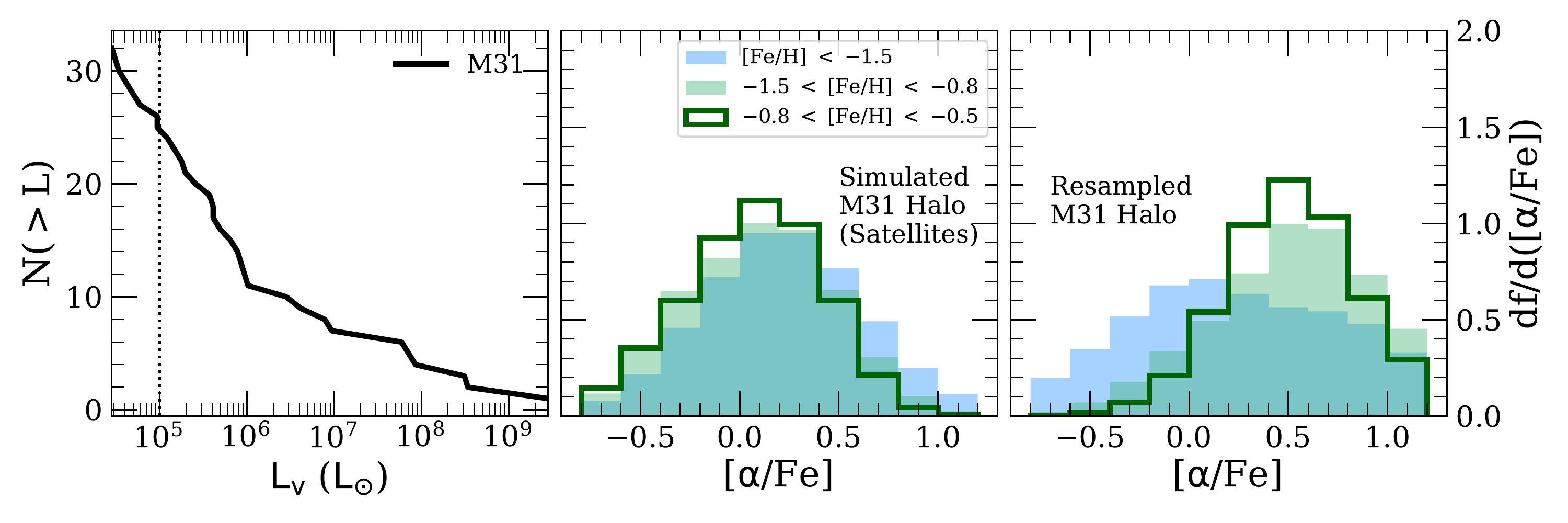}
    \caption{The construction of the inner stellar halo of M31 from present-day M31 satellite galaxies (\S~\ref{sec:halo_from_dwarfs}). ({\it Left panel}) V-band luminosity function of satellite galaxies within 300 kpc of M31, where absolute V-band magnitudes were taken from the compilation by \citet{McConnachie2012}. The dotted line represents the luminosity above which the luminosity function is likely to be complete ($L_V$ $>$ 10$^{5}$ $L_\odot$). The luminosity function is used to assign weights to the abundance distributions of M31 satellite galaxies contributing to the simulated stellar halo of M31 ({\it middle panel}). The simulated stellar halo is represented by \alphafe\ distribution functions, separated into three metallicity bins. ({\it Right panel}) The bootstrap re-sampled observed \alphafe\ distribution functions, separated into metallicity bins, of the stellar halo of M31, as probed by the stellar halo components in 5 M31 fields (this work, \paperii; $r_{\rm{proj}}$ $\lesssim$ 26 kpc). The smooth inner stellar halo of M31 is more $\alpha$-enhanced for \feh\ $\gtrsim$ $-$1.5 than would be expected for a stellar halo constructed from present-day M31 satellite galaxies.}
    \label{fig:m31_halo}
\end{figure*}

Numerous simulations have investigated stellar halo formation via accretion in the context of $\Lambda$CDM cosmology, where stellar halos of massive host galaxies are predicted to form hierarchically from smaller, disrupted stellar systems \citep{BullockJohnston2005,Font2006,Font2008,Font2011,Zolotov2009,Zolotov2010,Cooper2010,Tissera2013}. The chemical abundance distributions of the stellar halo of MW and M31-like galaxies should therefore reflect the properties of the constituent progenitor galaxies. Given that the \alphafe\ distribution at a given metallicity of the MW stellar halo disagrees with that of present-day MW dSphs \citep{Unavane1996,Shetrone2003,Venn2004}, we investigated whether the stellar halo of M31 could have formed from a population of progenitors similar to present-day M31 satellite galaxies.

To construct simulated abundance distributions for a M31 stellar halo formed from M31 satellite galaxies, we assumed that the progenitors in this scenario possessed a luminosity function equivalent to the luminosity function of satellite galaxies within 300 kpc of M31 (left panel of Figure~\ref{fig:m31_halo}), where properties for M31 satellite galaxies were taken from the compilation by \citet{McConnachie2012}. We utilized M31 satellites with existing \alphafe\ and \feh\ abundance measurements ($N$ $>$ 20) from \citet{Vargas2014a} (NGC 185, And II) and \replaced{Kirby et al., in preparation}{\citet{Kirby2019}} (And VII, And I, And III, And V), spanning $M_{\star}$ $\sim$ 10$^{5-7}$ $M_{\odot}$, based on deep DEIMOS 1200G spectra.\footnote{The median S/N of the \replaced{Kirby et al.}{\citeauthor{Kirby2019}} sample of dSphs is $\sim$23 \AA$^{-1}$, which is slightly higher than the stellar halo sample ($\sim$17 \AA$^{-1}$). The S/N of the \citeauthor{Vargas2014a} sample ranges from 15$-$25 \AA$^{-1}$. The measurement uncertainties on \feh\ are comparable between the combined \citeauthor{Vargas2014a} and \replaced{Kirby et al.}{\citeauthor{Kirby2019}} M31 satellite sample ($\delta$(\feh) $\sim$ 0.13, $\delta$(\alphafe) $\sim$ 0.23) and our M31 stellar halo sample ($\delta$(\feh) $\sim$ 0.14, $\delta$(\alphafe) $\sim$ 0.29). Thus, we anticipate that the bias from S/N limitations (\S~\ref{sec:sample_select}) similarly affects both samples.} Each individual RGB star, $i$, with measurements of \alphafe\ and \feh\ was assigned a probability, $p_{i,j}$, of contributing to the simulated stellar halo based on the stellar mass, $M_{\star, j}$, and V-band luminosity, $L_{V,j}$, of its host satellite galaxy,

\begin{equation}
p_{i,j} = \frac{ M_{\star,j} \Phi (L_{V,j}) / N_j }{ \Sigma_{j=1}^{N_{gal}} M_{\star,j} \Phi (L_{V,j})},
\label{eq:p_dsph}
\end{equation}
where $\Phi$ is the V-band luminosity function of present-day M31 satellite galaxies, $N_j$ is the number of RGB stars with abundance measurements in galaxy $j$, and $N_{gal}$ is the total number of M31 satellite galaxies contributing to the abundance distribution of the simulated stellar halo. We consider only the luminosity range over which the luminosity function is likely to be complete ($L_V$ $>$ 10$^{5}$ $L_\odot$), and only RGB stars with \feh\ $<$ $-$0.5 (\S~\ref{sec:dsph}) and $\delta$(\alphafe) $<$ 0.5 (\S~\ref{sec:sample}).

To construct the abundance distributions, we drew 10$^{6}$ random samples from the observed abundance distribution of M31 satellite galaxies ($N_{\rm{tot}}$ = 278) according to the probability distribution defined in Eq.~\ref{eq:p_dsph}. Additionally, we perturbed the observed abundance distribution during each draw by the uncertainties on the measurements, assuming Gaussian errors. Figure~\ref{fig:m31_halo} (middle panel) presents \alphafe\ distributions for the simulated stellar halo of M31 for a few metallicity bins. The \alphafe\ distributions for the high metallicity bins (\feh\ $>$ $-$1.5) are less $\alpha$-enhanced on average compared to the low metallicity bin (0.07$-$0.09 dex vs. 0.22 dex), reflecting the typical declining abundance pattern of \alphafe\ vs. \feh\ for present-day dwarf galaxies.

Figure~\ref{fig:m31_halo} also shows bootstrap re-sampled \alphafe\ distributions of the observed abundance distribution of M31's stellar halo ($r_{\rm{proj}}$ $\lesssim$ 26 kpc) for various metallicity bins. We constructed the abundance distributions based on abundances from the stellar halo components ($p$ $<$ 0.5; \S~\ref{sec:prob_substruct}) of the 5 total M31 fields presented in this work and \paperii\ ($N_{\rm{tot}}$ = 29), using the same criteria as in the case of the simulated stellar halo. The stellar halo of M31 is more $\alpha$-enhanced by 0.43$-$0.50 dex between $-$1.5 $<$ [Fe/H] $<$ $-$0.5 than expected for a stellar halo formed from progenitors with properties similar to those of present-day M31 satellites\footnote{The intermediate and high metallicity bins are statistically consistent with one another for the re-sampled stellar halo, although the high metallicity bin has a lower $\langle$\alphafe$\rangle$ by $\sim$0.08. The difference in the means may be a result of small sample sizes, or alternatively contamination in the stellar halo by substructure at [Fe/H] $>$ $-$0.8, owing to limitations of our kinematic decomposition (\S~\ref{sec:kinematic_decomp})}. Interestingly, $\langle$\alphafe$\rangle$ for the low metallicity bin (\feh\ $<$ $-$1.5) of the re-sampled stellar halo is nearly identical to that of the simulated stellar halo. Using two-sample KS tests, with $10^{4}$ draws of $N=29$ measurements from the parent stellar halo distributions, we find that the \alphafe\ distributions at high metallicity (\feh \ $>$ $-$1.5) are inconsistent between the re-sampled stellar halo and the simulated stellar halo at the $p$ $<$ 1\% level, whereas the low metallicity distributions are consistent.\footnote{Given that we compared \alphafe\ distributions in metallicity bins and consider only [Fe/H] $<$ $-$0.5, the bias against red, presumably metal-rich, stars affected by TiO absorption in the M31 stellar halo sample (\S~\ref{sec:sample},~\ref{sec:sample_select}) should not alter these conclusions.} 

Thus, based on currently available abundance measurements, we conclude that the metal-rich (\feh\ $>$ $-$1.5) inner stellar halo of M31 ($r_{\rm{proj}}$ $\lesssim$ 26 kpc) is unlikely to have formed from disrupted dwarf galaxies with properties similar to present-day M31 satellite galaxies. This is in agreement with findings that the global properties of M31's stellar halo are consistent with dominant contributions from massive progenitor(s) with $M_{\star}$ $\sim$ 10$^{8-9}$ $M_{\odot}$ \citep{Font2011,Deason2016,DSouzaBell2018a,Monachesi2019}.

\subsection{Inner Halo Substructure and Present-Day Satellite Galaxies}
\label{sec:dsph}

\begin{figure}
    \centering
    \includegraphics[width=\columnwidth]{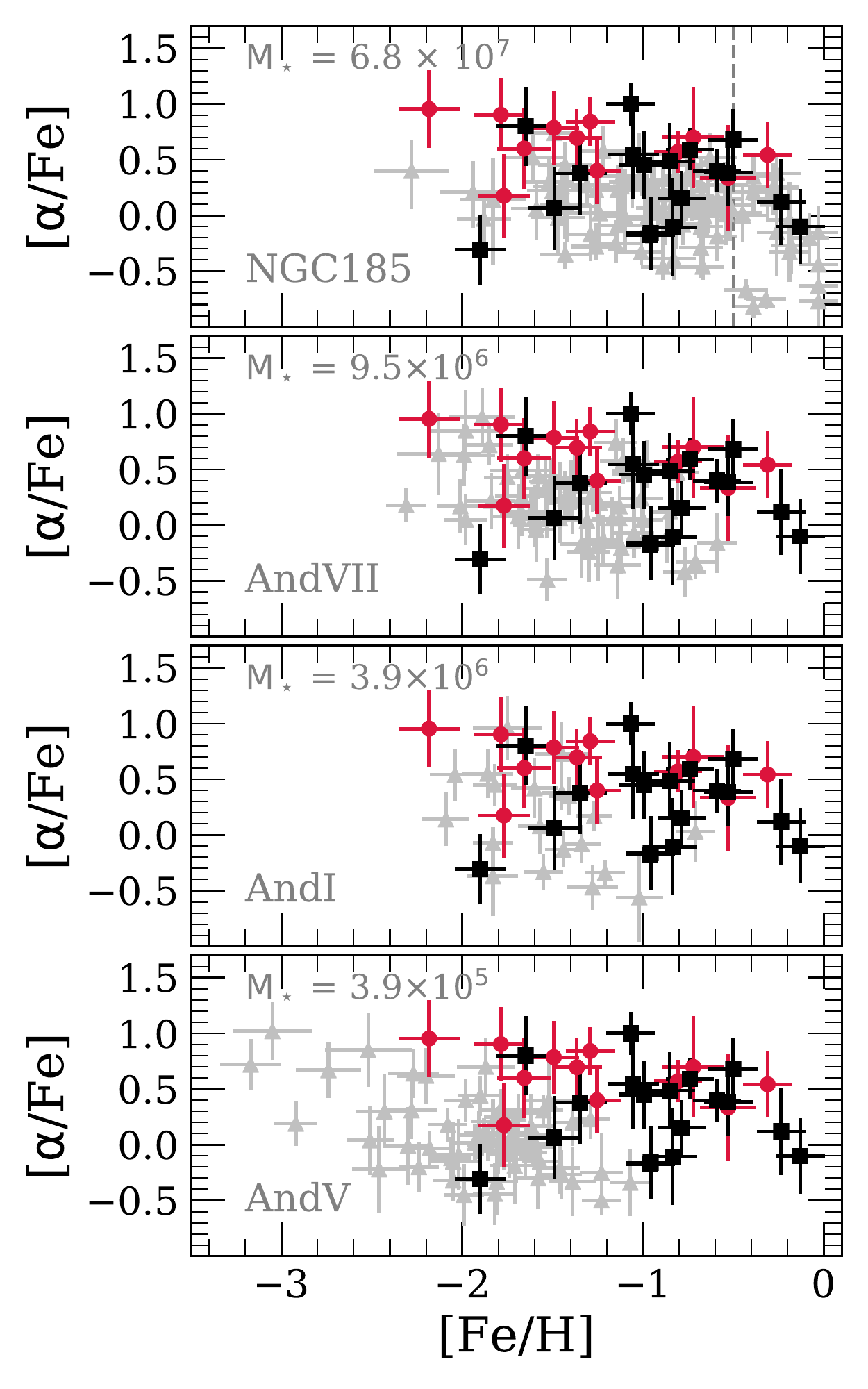}
    \caption{\alphafe\ vs. \feh\ for M31 RGB stars that are likely to belong to 12 kpc SE shelf feature (\textit{red circles}) and 22 kpc GSS core and KCC (\textit{black squares}) compared to the M31 satellite galaxies (\textit{grey triangles}; \S~\ref{sec:dsph}) NGC 185 \citep{Vargas2014a}, And VII, And I, and And V (Kirby et al., in prep). From top to bottom, the satellite galaxies are ordered according to decreasing luminosity, where stellar masses are adopted from \citet{McConnachie2012}. The vertical dashed line (\feh\ = $-$0.5) delineates the metallicity above which the \alphafe\ measurements of \citet{Vargas2014a} become uncertain. The \feh\ distributions of substructure in the 12 kpc and 22 kpc fields resemble satellite galaxies with $M_{\star}$ $\sim$ 10$^{6}$ M$_{\odot}$ and $M_{\star}$ $\gtrsim$ 10$^{7}$ $M_{\odot}$, respectively.}
    \label{fig:dsphs}
\end{figure}

The progenitor of the GSS is predicted to have been a massive dwarf galaxy of at least $M_{\star}$ $\sim$ 10$^{9}$ $M_\odot$ (e.g., \citealt{Fardal2006, MoriRich2008}), and therefore abundances in the GSS should in principle reflect abundance patterns characteristic of massive dwarf galaxies. If the SE shelf in fact originates from the GSS progenitor (\S~\ref{sec:se_shelf}), we might also expect its abundance distributions to match that of dwarf galaxies. Thus, we compare the metallicity and $\alpha$-element abundances of  substructure in the 12 kpc halo and 22 kpc GSS fields to a sample of M31 satellite dwarf galaxies with measured abundances (NGC 185 and And II from \citealt{Vargas2014a}; And VII, And I, And III, and And V, from \replaced{Kirby et al., in preparation}{\citet{Kirby2019}}. Figure~\ref{fig:dsphs} illustrates a subset of this comparison. We classified M31 RGB stars as belonging to substructure if they were more likely to be associated with substructure than the stellar halo (\S~\ref{sec:prob_substruct}). In the case of the GSS field, we do not distinguish between the GSS core and the KCC. 


Using a KS test, we find that the metallicity distribution of substructure in the 22 kpc GSS field is consistent with a dwarf galaxy at least as massive as NGC 185 ($M_{\star}$ = 6.8 $\times$ 10$^{7}$ $M_{\odot}$; \citealt{McConnachie2012}).\footnote{We considered only [Fe/H] $<$ $-$0.5 for the comparison between the abundances of the substructure components in H and S and NGC 185, owing to uncertainty in the abundances of NGC 185 above this metallicity \citep{Vargas2014a}.} 
Based on the mean metallicity of GSS abundances at 17 kpc ($-0.87$ $\pm$ 0.10 dex), \paperii\ used the stellar mass--metallicity relation for Local Group dwarf galaxies \citep{Kirby2013} to estimate that the GSS progenitor had a stellar mass of at least $\sim$0.5$-$2$\times$10$^{9}$ $M_{\odot}$. Given that the mean metallicity of the GSS at 22 kpc agrees with that at 17 kpc ($\langle$\feh$\rangle_{\rm{GSS,22kpc}}-\langle$\feh$\rangle_{\rm{GSS,17kpc}}$ = $-$0.15 $\pm$ 0.17), our results corroborate the GSS progenitor mass inferred by \paperii, where both samples are similarly biased against red stars (\S~\ref{sec:sample_select}).

Stars in both the 17 kpc and 22 kpc GSS fields are more $\alpha$-enhanced than NGC 185. \paperii\ found that the $\alpha$-element abundances of the GSS at 17 kpc were similarly $\alpha$-enhanced compared to Sagittarius, the Large Magellanic Cloud, and the Small Magellanic Cloud, where these conclusions also apply to the GSS at 22 kpc (Figure~\ref{fig:gss_S_f207}). The $\alpha$-element abundances of the GSS at 17 kpc and 22 kpc imply that the GSS progenitor experienced a higher star formation efficiency than NGC 185. Based on {\it HST} imaging, NGC 185 shows evidence for recent and extended star formation within its inner 200 pc \citep{ButlerMartinez-Delgado2005,Weisz2014}, quenching $\sim$3 Gyr ago. The {\it HST} CMD-based SFH for the GSS field (Table~\ref{tab:m31_fields}) implies that star formation ceased in the GSS progenitor $\sim$4-5 Gyr ago \citep{Brown2006}, presumably when interactions with M31 began to affect its evolution. Thus, although the GSS progenitor may have quenched $\sim$1-2 Gyr earlier than NGC 185, the galaxy had reached at least the same metallicity by that epoch, further supporting the hypothesis of a comparatively high star formation efficiency for the GSS progenitor. 

Although the \alphafe\ distributions of the GSS fields and NGC 185 differ, they have a similar metallicity spread. NGC 185 possesses a negative radial metallicity gradient out to $\sim$2.2 kpc \citep{Vargas2014a}, assuming $d_{\odot}$ = 617 kpc \citep{McConnachie2005} and $r_{h}$ = 1.5' \citep{deRijcke2006}, although its stellar mass is significantly lower than the inferred mass of the GSS progenitor. In accordance with expectations (e.g., \citealt{Fardal2008}), the GSS progenitor may have had a \added{radial} metallicity gradient. If so, the abundances of the 17 kpc and 22 kpc GSS fields may probe stellar populations from a large radial range in the progenitor (\paperii; \citealt{Hammer2018}).


Interestingly, the 22 kpc GSS field possesses an \alphafe\ distribution that is statistically consistent with that of satellite galaxies with $M_{\star}$ $\sim$ 0.83$-$9.5 $\times$ 10$^{6}$ M$_{\odot}$,
although the metallicity distribution of the substructure is incompatible with that of the lower mass ($M_{\star}$ $<$ 10$^{7}$ $M_{\odot}$) dwarf galaxies. These lower mass dwarf galaxies had relatively truncated SFHs, forming at least 50\% of their stellar mass as of 10 Gyr ago \citep{Weisz2014,Skillman2017}. 
This may indicate that stars in the GSS core, KCC, and lower mass dwarf galaxies may have similar contributions of core-collapse supernovae relative to Type Ia supernovae, with the caveat that the GSS progenitor likely experienced a higher star formation efficiency and extended SFH compared to these systems.


The metallicity distribution of the SE shelf resembles that of satellite galaxies with 3.9$-$9.5 $\times$ 10$^{6}$ $M_{\odot}$,
although its $\alpha$-element distribution is inconsistent with the sample of M31 satellite galaxies across the entire analyzed mass range. The implications of this comparison 
are less straightforward, particularly considering the bias against red stars in the SE shelf (\S~\ref{sec:sample_select}, \S~\ref{sec:abund_substruct}) and the possibility of contamination of the SE shelf sample by halo stars.
If the SE shelf abundances are representative, the SE shelf could originate from a progenitor galaxy with $M_\star \sim$ 10$^{6-7} M_\odot$, which possessed relatively short Type Ia supernovae timescales compared to present-day satellites of similar mass, that is distinct from the GSS progenitor. Alternatively, the GSS progenitor could have possessed a significant \added{radial} metallicity gradient, such that SE shelf originates from a chemically distinct region of the GSS progenitor. We further evaluate these possibilities in \S~\ref{sec:se_shelf}.

\subsection{Is the SE Shelf Related to the GSS Progenitor?}
\label{sec:se_shelf}

\begin{figure}
    \centering
    \includegraphics[width=\columnwidth]{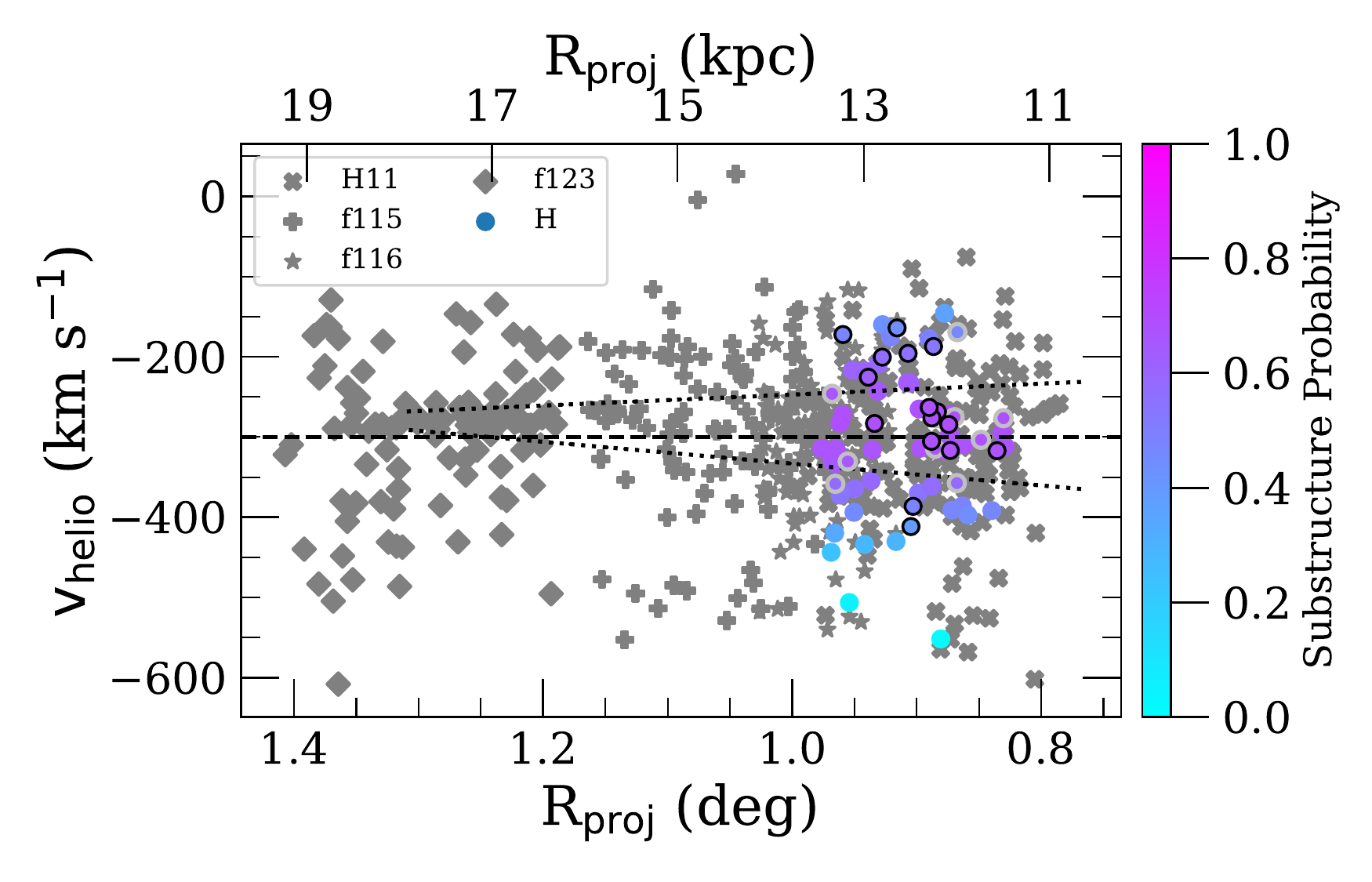}
    \caption{Heliocentric velocity versus projected distance of M31 RGB stars (\S~\ref{sec:se_shelf}). The 12 kpc field (H) corresponds to circles color-coded according to probability of belonging to substructure (\S~\ref{sec:prob_substruct}), where M31 RGB stars in our final abundance sample (\S~\ref{sec:sample}) are outlined in black and TiO stars are outlined in light grey. Dark grey points represent various DEIMOS fields with shallow 1200G spectroscopy that show evidence of the Southeast shelf (Figure~\ref{fig:fields}; \citealt{Gilbert2007}, the predicted shell formed from GSS progenitor stars on their fourth pericentric passage \citep{Fardal2007}. The dashed horizontal line corresponds to M31's systemic velocity, whereas the dotted lines correspond to the observed boundaries of the Southeast shelf in this space \citep{Gilbert2007}. The substructure in the 12 kpc field fits within the spatial and kinematical profile of the Southeast shelf.}
    \label{fig:se_shelf}
\end{figure}

The inner halo of M31 contains abundant substructure, most of which is likely associated with the extended disk or the GSS merger event (e.g., \citealt{Ferguson2005,Ibata2007,McConnachie2018}). In particular, \citet{Gilbert2007} identified a kinematically cold feature at $-$300 \kms\
using spectroscopy of $\sim$1000 M31 RGB stars between 9$-$30 kpc in M31's southeastern quadrant. The velocity dispersion of the feature decreased with increasing projected radial distance, from $\sigma_v$ = 56 \kms\ at 12 kpc to $\sigma_v$ = 10 \kms\ at 18 kpc, reflecting the characteristic pattern of a shell system originating from a disrupted progenitor galaxy. Based on its spatial and kinematic properties, \citet{Gilbert2007} associated the $-$300 \kms\ cold component with the SE shelf, a predicted, faint shell corresponding to the fourth pericentric passage of GSS progenitor stars \citep{Fardal2007}.

The 12 kpc field overlaps with DEIMOS fields (Figure~\ref{fig:fields}) in which \citet{Gilbert2007} identified the SE shelf. The velocity dispersion of the 12 kpc substructure ($\sigma_v$ = 66 \kms; Table~\ref{tab:kinematic_decomp}) is similar to that of the SE shelf at the same radius. Figure~\ref{fig:se_shelf} shows the heliocentric velocity versus the radial projected distance of the 12 kpc field compared to M31 RGB stars in DEIMOS fields with shallow 1200G spectroscopy, where \citet{Gilbert2007} identified the fields as contributing to the SE shelf. The M31 RGB stars that are most likely to belong to substructure in the 12 kpc field fall within the observed spatial and kinematical profile of the SE shelf \citep{Gilbert2007}. Thus, based on these properties alone, the 12 kpc field is likely polluted by material from the GSS progenitor.

The properties of the stellar population in the vicinity of the 12 kpc field also argue in favor of its contamination by GSS progenitor stars. \citet{Brown2006} and \citet{Richardson2008} found that the stellar age and photometric metallicity distributions in the {\it HST}/ACS halo11 field (Figure~\ref{fig:fields}, Table~\ref{tab:m31_fields}), which overlaps with the 12 kpc field, and the {\it HST}/ACS stream field were remarkably similar. Additionally, \citet{Gilbert2007} observed that \fehphot\ was similar between M31 RGB stars likely belonging to the $-$300 \kms\ cold component and the GSS. 

If the 12 kpc substructure corresponds to the SE shelf, it may differ from the mean metallicity of the GSS core by $-$0.28$^{+0.20}_{-0.18}$ dex (Table~\ref{tab:m31_avg_abund}). This quoted value is weighted by the probability of belonging to kinematic substructure for all stars in the field.  However, the maximum substructure probability is low (69\%). In other words, M31 RGB stars with kinematic properties matching that of the SE shelf (with $p >$ 0.5; \S~\ref{sec:prob_substruct}) have a 35\% chance on average of belonging to the stellar halo. 

If the quoted metallicity difference between the SE shelf feature and the GSS core is accurate, this could indicate that the SE shelf originated from a chemically distinct region of the GSS progenitor. Although no metallicity gradient has been observed along the GSS, there is evidence of a gradient between the GSS core and its outer envelope \citep{Ibata2007, Gilbert2009b}, such that GSS formation models have explored the possibility of the observed metallicity gradient originating from a \added{radial} gradient in the GSS progenitor \citep{Fardal2008,Miki2016,Kirihara2017}.

\subsection{Abundances in the Outer Disk of M31 and the MW}
\label{sec:disk_abund}

Few studies of the metallicity of stars in M31's outer disk exist in the literature. \citet{Collins2011} measured \ion{Ca}{2}-triplet based \feh\ for 21 DEIMOS fields between 10-40 projected kpc on the sky from M31's center along the southwestern major axis of M31, finding that 
$\langle$[Fe/H]$\rangle_{\rm{CaT, thin}}$ = $-$0.7 and $\langle$[Fe/H]$\rangle_{\rm{CaT, thick}}$ = $-$1.0, where the thin disk has an average velocity dispersion of 36 \kms\ vs. 51 \kms\ for the thick disk. Thus, both the metallicity ($\langle$[Fe/H]$\rangle$ = $-$0.82) and velocity dispersion ($\sigma_v$ = 16 \kms) of the 26 kpc disk suggest it is similar to M31's thin disk, or potentially the extended disk of M31 (\S~\ref{sec:disk_formation}).

 In the inner disk of M31, \citet{Gregersen2015} constructed photometric stellar metallicity distributions, assuming constant stellar age and \alphafe\ = 0, based on 7 million RGB stars across the PHAT \citep{Dalcanton2012} footprint in M31's northeastern disk. They found a \added{radial} metallicity gradient of $-$0.020 dex kpc$^{-1}$ between $r_{\rm{disk}}$ $\sim$4$-$20 kpc, with a \fehphot\ normalization of $\sim$0.11. Extrapolating this metallicity gradient, we estimated that \fehphot\ would be $-$0.6 at the location of 26 kpc disk field, $r_{\rm{disk}}$ = 35 kpc. In comparison, we calculated \fehphot\ = $-$0.88 for our isolated disk feature in this field. We caution that the behavior of the \added{radial} metallicity gradient from individual RGB stars in M31's disk is unknown at large radii \citep{Kwitter2012, Sick2014}, and that differences in metallicity measurement methodology will impact the absolute metallicity normalization.

The dearth of chemical abundance data in the outer disk of M31 applies to the MW as well. However, the stellar metallicity distribution in the MW disk has been well-studied through spectroscopic surveys out to $\sim$15 kpc, finding a comparatively steep \added{radial} metallicity gradient of $\sim-$0.06 dex kpc$^{-1}$ (e.g., \citealt{Cheng2012a, Hayden2014, Boeche2014, Mikolaitis2014}). In particular, using $\sim$70,000 RGB stars from APOGEE, \citet{Hayden2014} found \feh\ = $-$0.43 in the MW disk plane between 13$-$15 kpc. If we perform the same exercise as in the case of M31's disk and extrapolate the MW's metallicity gradient to $r_{\rm{disk}}$ = 35 kpc, we would obtain \feh\ $\sim-$0.8, which is similar to our measured mean metallicity in the 26 kpc M31 disk field.

Interestingly, spectroscopic abundances exist for the Triangulum-Andromeda (TriAnd) overdensity \citep{Majewski2004,Rocha-Pinto2004}, a distant structure ($r_{\rm{GC}}$ $\sim$ 20 kpc) potentially associated with an extension of the MW disk \citep{Price-Whelan2015, Xu2015, Li2017}. Recently, \citet{Hayes2018b} found $\langle$[Fe/H]$\rangle$ = $-$0.8 for TriAnd, in agreement with our measured metallicity for M31's 26 kpc disk. Additionally, TriAnd has chemical abundances (including $\alpha$-element abundances) similar to the most metal-poor stars in the MW's ``outer disk'' ($r_{\rm{GC}}$ $>$ 9 kpc) \citep{Hayes2018b, Bergemann2018}.  \added{The outer disk of M31 may be chemically the most similar to a potential third component of the MW's disk, known as the metal-weak thick disk \citep{Carollo2019}, which is metal-poor (\feh\ $\sim$ $-$1) and relatively $\alpha$-enhanced (\alphafe\ $\gtrsim$ 0.22). However, evidence for this component has thus far only been detected in the solar neighborhood, and its kinematics ($\sigma_v$ $\sim$ 60 \kms) are inconsistent with those of M31's disk at 26 kpc.}

Given that this work presents the first \alphafe\ measurements in M31's disk, we are limited to the MW's disk for comparisons of \alphafe\ measurements based on individual stars. The MW disk is known to possess high-$\alpha$ and low-$\alpha$ sequences at subsolar \feh\ \citep{Bensby2011a, Adibekyan2013, Nidever2014}. High-$\alpha$ stars have been associated with the MW's thick disk (e.g., \citealt{Bensby2005,Reddy2006, Lee2011b}) and have ages exceeding $\sim$7 Gyr, where the most $\alpha$-enhanced stars tend toward older ages \citep{Bensby2005, Haywood2013}. However, a population of young $\alpha$-rich stars has also been identified in the MW disk \citep{Martig2015, Chiappini2015}. In this instance, high-$\alpha$ refers to \alphafe\ $\sim$ 0.3, which is significantly lower than our measurement of \alphafe\ $\sim$ 0.6 for M31's disk (and M31's stellar halo) at $r_{\rm{disk}}$ = 35 kpc. If the mean stellar age of stars in the 26 kpc disk field is $\sim$7 Gyr \citep{Brown2006,Bernard2015a}, with a negligible population of stars with ages $\gtrsim$10 Gyr, then it is similar in age to, if not younger than, the MW disk's high-$\alpha$ population. Assuming that the 26 kpc disk feature is representative of M31's outer disk, the expected discrepancy in \alphafe\ between the MW and M31's outer disk is potentially even larger, considering that the low-$\alpha$ sequence is more prominent in the outer disk of the MW \citep{Cheng2012b, Bovy2012, Nidever2014, Hayden2015}. 

\subsection{Disk Formation Scenarios: the MW vs. M31}
\label{sec:disk_formation}

The patterns of \alphafe\ vs.\ \feh\ in the MW disk with respect to scale length and scale height  \citep{Bensby2011a,Cheng2012b,Bovy2012,Anders2014,Nidever2014,Hayden2015} provide support for scenarios in which the inner disk of the MW formed prior to the outer disk (and the chemical thick disk was formed before the chemical thin disk). In particular, the dominance of the low-$\alpha$ sequence in the outer disk and the homogeneity of the high-$\alpha$ sequence in the inner disk (where the scale length is $\sim$2 kpc) could result from a combination of an initial stellar population that formed from a gas-rich, well-mixed, turbulent interstellar medium and multiple distinct stellar populations in the outer disk \citep{Nidever2014}. These outer disk populations could result from a transition from low- to high- star formation efficiency coupled with extended pristine gas infall \citep{Chiappini2001} or increasing outflow rate with increasing galactocentric radius. Based on the chemical abundance patterns of the MW inner vs. outer disk \citep{Bovy2012,Nidever2014}, radial migration appears to have played a significant role in the evolution of the MW's disk (e.g., \citealt{SellwoodBinney2002,SchonrichBinney2009a}), although its efficiency must have been limited to match the lack of observed high-$\alpha$ stars in the outer disk \citep{Cheng2012b}. In general, the abundance patterns of the MW disk seem to disfavor an external origin (e.g., \citealt{Brook2012,Minchev2014}), although this possibility cannot be excluded (see also \citealt{Mackereth2019b}).

The fact that the outer disk of M31 is $\alpha$-enhanced relative to the MW disk between $\sim$9-15 kpc suggests that M31's outer disk may have experienced a different formation history or internal evolution. Marked differences in the structural morphology \citep{Ibata2005} and dynamics \citep{Dorman2015, Quirk2019} of M31's disk support this hypothesis. Perhaps the most distinguishing feature of M31's disk relative to the MW is its ubiquitous burst of star formation that occured 2$-$4 Gyr ago \citep{Bernard2015a, Bernard2015b, Williams2015, Williams2017}. Taking into account the unusual SFH of M31's disk, coupled with its relatively large velocity dispersion and steep age--velocity dispersion correlation \citep{Dorman2015}, \citet{Hammer2018} found that the observed properties of M31's disk are consistent with a 4:1 major merger, in which first passage occurred 7$-$10 Gyr ago and nuclei coalescence occured 2$-$3 Gyr ago. 

Possible origins for the extended disk of M31 ($r$ $\lesssim$ 40 kpc) are the accretion of multiple small systems or a single secondary progenitor \citep{Ibata2005}. An episode of star formation induced by a major merger offers the advantage of explaining both the disk-like kinematics and chemical homogeneity of the extended disk ([Fe/H]$_{\rm CaT}$ = $-$0.9). The low velocity dispersion of the 26 kpc disk ($\sigma_v$ $\sim$ 16 \kms), its high $\alpha$-element abundance ($\langle$\alphafe$\rangle$ = 0.58), and relatively young stellar age (4$-$8 Gyr old; \citealt{Brown2006}) are consistent with an extended disk that experienced rapid star formation induced by a major merger. The accretion of multiple small systems along the plane of the disk is less likely to result in such a high $\alpha$-element abundance, presuming that such systems would relatively chemically evolved, and thus more $\alpha$-poor. Based on the relatively young stellar age of the disk compared to the 23 kpc field ($\sim$7.5 Gyr in the disk field vs. 10$-$11 Gyr in the 23 kpc field; \citealt{Brown2006,Brown2007}), the expectation from the accretion of small systems would be that the 23 kpc field is more $\alpha$-enhanced than the disk, in contradiction to our abundance measurements for these fields. Furthermore, the chemical abundances of the GSS are consistent with a major merger scenario (as in \citealt{Hammer2018} or \citealt{DSouzaBell2018b}), assuming that the stars in the GSS core do not predominantly originate from the center of the progenitor and the GSS progenitor had a \added{radial} metallicity gradient (\paperii).

Internal mechanisms, such as radial migration, are problematic in terms of explaining the $\alpha$-enhancement of M31's disk at 26 kpc. This scenario requires the $\alpha$-enhanced population in the outer disk to have originated from an old, centrally concentrated stellar population, whereas the 26 kpc field contains a significant population of young stars \citep{Brown2006}. Additionally, the velocity dispersion of M31's disk is larger than that of the MW \citep{Dorman2015}, where the efficiency of radial migration is expected to decrease with increasing velocity dispersion \citep{Solway2012}. In light of current observations of M31's disk, we find that star formation induced by a major merger provides the simplest explanation for the chemical abundances of the 26 kpc disk.


\section{Conclusions}
\label{sec:summary}

We measured \alphafe\ and \feh\ from deep, low-resolution DEIMOS spectroscopy of 70 M31 RGB stars across the inner halo, Giant Stellar Stream \citep{Ibata2001}, and outer disk of M31. This is the largest detailed abundance sample in M31 to date, and in combination with \citet{Escala2019} and \replaced{Gilbert et al., submitted}{\citet{Gilbert2019}}, presents the first measurements of spectroscopic \feh\ and \alphafe\ in the inner halo, GSS, and disk of M31. Using a kinematic decomposition, we separated the stellar populations in our spectroscopic fields into ``smooth'' stellar halo and substructure. The substructure identified at 12 kpc, 22 kpc, and 26 kpc correspond to the Southeast shelf \citep{Fardal2007,Gilbert2007}, the GSS core and KCC \citep{Kalirai2006a,Gilbert2009b,Gilbert2019} and M31's outer disk, respectively. We summarize our primary results below.

\begin{enumerate}
    \item The inner halo, GSS, and outer disk of M31 are $\alpha$-enhanced ($\langle$\alphafe$\rangle$ $\gtrsim$ 0.35), where the 26 kpc disk and  22 kpc GSS fields are more metal-rich 
    than the 12 and 23 kpc inner halo fields 
    ($\Delta$(\feh)$_{\rm{26-12kpc}}$ = 0.38 $\pm$ 0.16, $\Delta$(\feh)$_{\rm{26-23kpc}}$ = 0.62$^{+0.17}_{-0.18}$, $\Delta$(\feh)$_{\rm{22-12kpc}}$ = 0.46 $\pm$ 0.15, $\Delta$(\feh)$_{\rm{22-23kpc}}$ = 0.70 $\pm$ 0.17).
    \item Measurements of \feh\ and \alphafe\ between 17 kpc (\paperii) to 22 kpc along the GSS are fully consistent. This is in agreement with previous studies illustrating the absence of a metallicity gradient along the GSS \citep{Ibata2007,Gilbert2009b}.
    \item 
    The inner halo of M31 ($r_{\rm{proj}}$ $\lesssim$ 26 kpc) appears to be more $\alpha$-enhanced than the MW inner halo at all radii. Additionally, we find suggestions that the outer halo of M31 \citep{Vargas2014b} is more $\alpha$-poor than the inner halo, although more data are necessary to confirm such a gradient. If a negative \added{radial} \alphafe\ gradient is present, it would agree with the implications of the steep negative \added{radial} \feh\ gradient of M31 \citep{Gilbert2014}, providing support for different progenitor(s) and/or formation mechanisms contributing to the inner versus outer halo.
    \item Based on currently available data, the \alphafe\ distribution of the metal-rich (\feh\ $>$ $-$1.5) inner stellar halo ($r_{\rm{proj}}$ $\lesssim$ 26 kpc) of M31 (i.e., with substructure removed) is inconsistent with having formed from the disruption of progenitors with chemical properties similar to present-day M31 satellite galaxies ($M_{\star}$ $\sim$ 10$^{5-7}$ $M_\odot$).
    \item In agreement with \paperii, comparisons to the abundance distributions of M31 satellite galaxies \citep{Vargas2014a,Kirby2019} suggest that the chemical properties of the GSS are consistent with a massive progenitor ($\gtrsim$ 0.5$-$2$\times$10$^{9}$ M$_{\odot}$; \paperii) that experienced a high star formation efficiency. Such comparisons also point to the SE shelf resembling lower mass dwarf galaxies ($M_\star$ $\sim$ 10$^{6}$ $M_\odot$), with the caveat of bias against red stars and potential stellar halo contamination in the SE shelf sample.
    \item  We found tentative evidence that the SE shelf is more metal poor than the GSS by $\gtrsim$ 0.10 dex, taking into account observational uncertainty. If the SE shelf in fact originates from the GSS progenitor \citep{Fardal2007,Gilbert2007}, then a \added{radial} metallicity gradient in the GSS progenitor (e.g., \citealt{Fardal2008}) could explain the observed metallicity difference.
    \item M31's disk at $r_{\rm{proj}}$ = 26 kpc ($r_{\rm{disk}}$ = 35 kpc) is consistent with nearly circular rotation \citep{Guhathakurta1988}, with $v_{\rm{lag}}$ = $-$9$^{+11}_{-3}$ \kms, and is dynamically cold ($\sigma_v$ = 16 \kms). The disk is highly $\alpha$-enhanced (\alphafe\ = 0.58) compared to the high-$\alpha$ population of the MW's disk (\alphafe\ $\sim$ 0.30). The metallicities of stars in the 26 kpc disk feature (\feh\ = $-$0.82) agree with predictions at comparable radii in the MW (based on extrapolation of its \added{radial} metallicity gradient, e.g., \citealt{Cheng2012a,Hayden2014}) and distant, possibly disk-related structures such as TriAnd \citep{Bergemann2018,Hayes2018b}.
    \item Taking into account the observed structural and dynamical properties of M31's disk \citep{Ibata2005,Dorman2015}, we find that a global episode of active star formation induced by a major merger \citep{Hammer2018,DSouzaBell2018b} is the simplest explanation for the observed chemical abundances of M31's disk at 26 kpc, \added{assuming that our sample is representative of this region}.
\end{enumerate}

Future work will continue to increase the sample size of M31 RGB stars with abundance measurements, such that we can place more stringent constraints on the accretion history of M31 and the formation of its stellar disk and halo.

\acknowledgments

\added{The authors thank the anonymous reviewer for a thorough reading of this manuscript and helpful comments.} \replaced{The authors}{We also} thank Stephen Gwyn for reducing the photometry for slitmasks H, S, and D and Jason Kalirai for the reductions of f130\_2. I.E. acknowledges support from a National Science Foundation (NSF) Graduate Research Fellowship under Grant No.\ DGE-1745301.  This material is based upon work supported by the NSF under Grants No.\ AST-1614081 (E.N.K.), AST-1614569 (K.M.G, J.W.), and AST-1412648 (P.G.).  E.N.K gratefully acknowledges support from a Cottrell Scholar award administered by the Research Corporation for Science Advancement, as well as funding from generous donors to the California Institute of Technology. E.C.C. was
supported by an NSF Graduate Research Fellowship and an
ARCS Foundation Fellowship, as well as NSF Grant AST-1616540. The analysis
pipeline used to reduce the DEIMOS data was developed at UC Berkeley with support from NSF grant AST-
0071048.

We are grateful to the many people who have worked to make the Keck Telescope and its instruments a reality and to operate and maintain the Keck Observatory.  The authors wish to extend special thanks to those of Hawaiian ancestry on whose sacred mountain we are privileged to be guests.  Without their generous hospitality, none of the observations presented herein would have been possible.

\vspace{5mm}
\facilities{Keck II/DEIMOS}
\software{astropy \citep{Astropy2013, Astropy2018}, emcee \citep{Foreman-Mackey2013}.}

\appendix

\section{Comparison Between DEIMOS 600ZD and 1200G Elemental Abundances}
\label{sec:600_vs_1200}

\begin{figure*}
    \centering
    \includegraphics[width=0.8\textwidth]{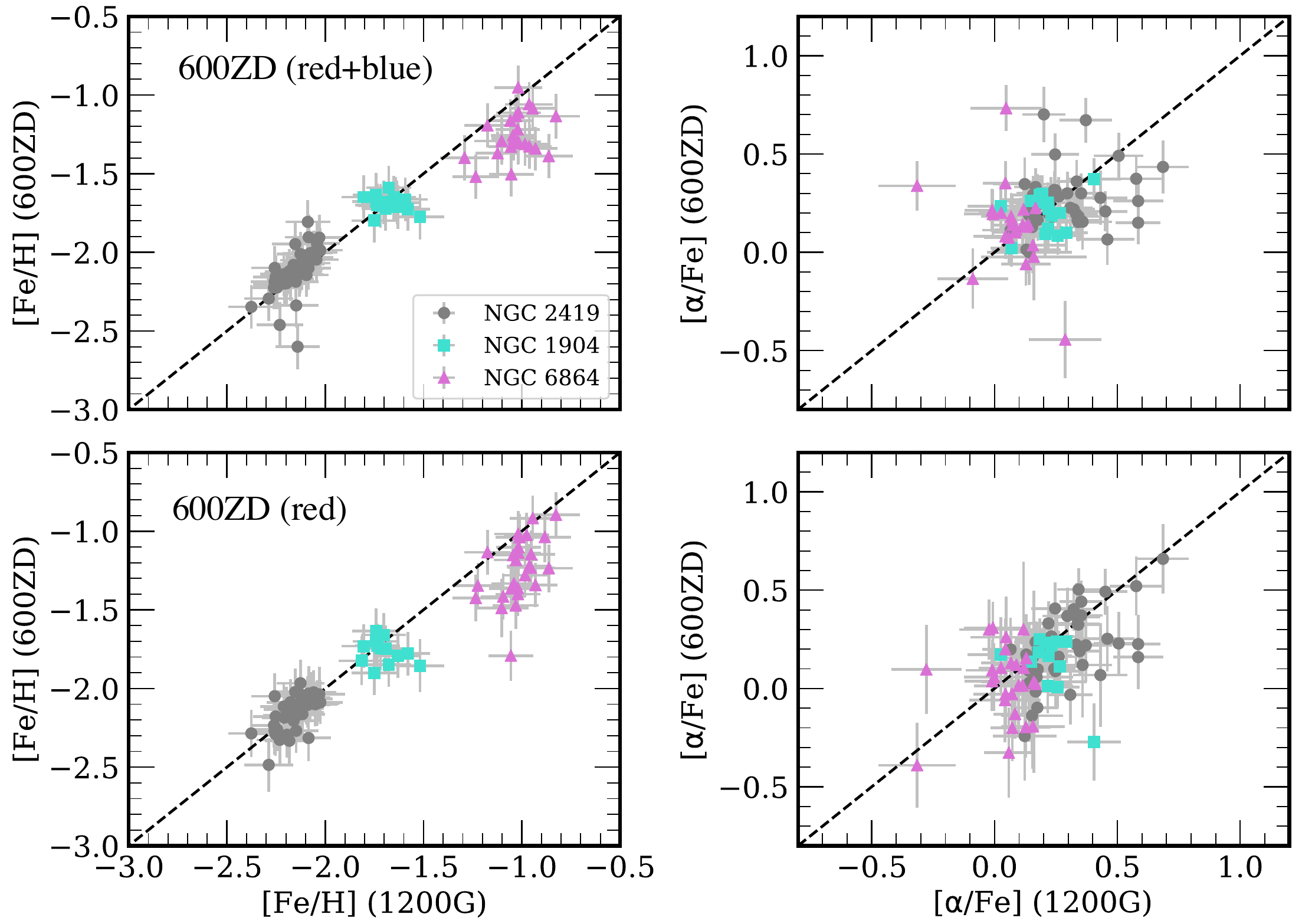}
    \caption{A star-by-star comparison between \feh\ and \alphafe\ measurements for giant stars from a sample of MW globular clusters \citep{Kirby2016, Escala2019} using spectra obtained with the 1200G and 600ZD gratings on DEIMOS. The dashed line represents a one-to-one relation. ({\it Top panels})
    Standard 600ZD-based abundance measurements ($N$ = 81) using the entire usable wavelength range (4500$-$9100 \AA) are broadly consistent (at 1.2$-$1.3$\sigma$) with 1200G-based abundances within the uncertainties. ({\it Bottom panels}) Restricting the abundance measurement for 600ZD spectra to only red wavelengths (6300$-$9100 \AA) reduces the scatter between the two techniques (to 0.9$\sigma$, $N = 84$), indicating that any excess scatter is owing to the inclusion of bluer wavelengths (4500-6300 \AA).
    }
    \label{fig:gcs}
\end{figure*}

\added{In \S~\ref{sec:abund_S}, \S~\ref{sec:abund_gradient}, and ~\S~\ref{sec:halo_from_dwarfs}, we simultaneously utilized measurements of \feh\ and \alphafe\ derived from the 1200G (\citealt{Vargas2014a}, Kirby et al. 2019, in press) and 600ZD (this work) gratings on DEIMOS. The spectral synthesis methods employed in each case \citep{Kirby2008,Kirby2009,Escala2019} are the same in principle, but rely on spectra of differing resolution ($R$ $\sim$ 6000 vs. 2500) and wavelength coverage (6300$-$ vs. 4500$-$9100 \AA). 

In this section, we illustrate the general consistency of the two measurement techniques using a sample of individual giant stars in MW globular clusters (GCs; NGC 2419, NGC 1904, NGC 6864) and MW dSphs (Draco, Canes Venatici I) observed with both the 1200G and 600ZD gratings. Measurements of \feh\ and \alphafe\ for the GCs were determined using 600ZD spectra by \paperi\ and using 1200G spectra by \citet{Kirby2016}, where identical slitmasks were used for each GC\@. \paperi\ presented measurements of a limited sample of stars from the 600ZD observations of  dSphs.  The measurements were restricted to those stars with previous abundance measurements in the literature.  In order to build a larger sample to compare results from 600ZD with 1200G, we measured abundances for the complete sample of 600ZD dSph stars in this work. The 1200G dSph abundance measurements are drawn from stars identified as members in the catalog of \citet{Kirby2010}. The photometry is identical for each star in common between 1200G and 600ZD measurements, where \teffphot\ and \loggphot\ are taken as inputs into the spectral synthesis software \citep{Kirby2008,Escala2019}. We refine our sample selection according to the relevant criteria outlined in \S~\ref{sec:sample_select}.

Figure~\ref{fig:gcs} shows a star-by-star comparison of \feh\ and \alphafe\ measured from both 1200G and 600ZD spectra for the GC sample. We do not find evidence of a statistically significant (to 3$\sigma$) error-weighted mean offset in either \feh\ or \alphafe\ between the 1200G and 600ZD measurements for this sample ($\langle\Delta$[Fe/H]$_{\rm 1200G-600ZD}\rangle$ = 0.0 $\pm$ 0.02, $\langle\Delta$[$\alpha$/Fe]$_{\rm 1200G-600ZD}\rangle$ = $-$0.08 $\pm$ 0.03). We quantify the consistency between the two measurement techniques by computing the standard deviation of their error-weighted difference,

\begin{equation}
    \sigma_\epsilon =  {\rm stddev} \left( \frac{ \epsilon_{\rm 1200G} - \epsilon_{\rm 600ZD} }{ (\delta\epsilon_{\rm 1200G})^{2} + (\delta\epsilon_{\rm 600ZD})^{2} } \right),
\end{equation}
where $\epsilon$ is a given chemical abundance measurement (\feh\ or \alphafe) and $\delta\epsilon$ is the associated measurement uncertainty. We omitted outliers in this calculation, performing iterative 5$\sigma$ clipping on the discrepancy between the 1200G and 600ZD abundances until the comparison sample converged. We find that $\sigma_{\rm [Fe/H]} = 1.15$ and $\sigma_{[\alpha/{\rm Fe}]} = 1.34$ for the GC sample, which indicates that the scatter between the 1200G and 600ZD measurements is not completely accounted for by the uncertainties ($\sigma_\epsilon = 1$). 

To investigate the source of this excess scatter, we re-measured abundances from the 600ZD spectra for the GCs, but we restricted the chemical abundance analysis to the same wavelength range used for 1200G spectra (6300$-$9100 \AA). Figure~\ref{fig:gcs} also illustrates the results of a star-by-star comparison between 1200G measurements and 600ZD measurements restricted only to redder wavelengths. In this case, the excess scatter disappears ($\sigma_{\rm [Fe/H]} = 0.86,  \sigma_{[\alpha/{\rm Fe}]} = 0.93$), where the measurements are consistent within the uncertainties. This indicates that the source of the excess scatter when utilizing the full wavelength range of the 600ZD spectra is the inclusion of bluer wavelengths (4500$-$6300 \AA), as opposed to lower spectral resolution. Thus, the scatter between 1200G- and 600ZD-based measurements is likely driven by the additional information contained in absorption features between 4500$-$6300 \AA\@.  For example, the bluer spectral regions contain a different balance of $\alpha$ absorption lines, such as the Mg~b triplet, than the red side.  Therefore the meaning of ``$\alpha$'' is different depending on the spectral range.  (This explanation does not apply to \feh.)  Alternatively, it is possible that the bluer regions of the spectrum have higher continuum normalization errors owing to the high density of absorption features at these wavelengths.} 

\begin{figure*}
    \centering
    \includegraphics[width=0.8\textwidth]{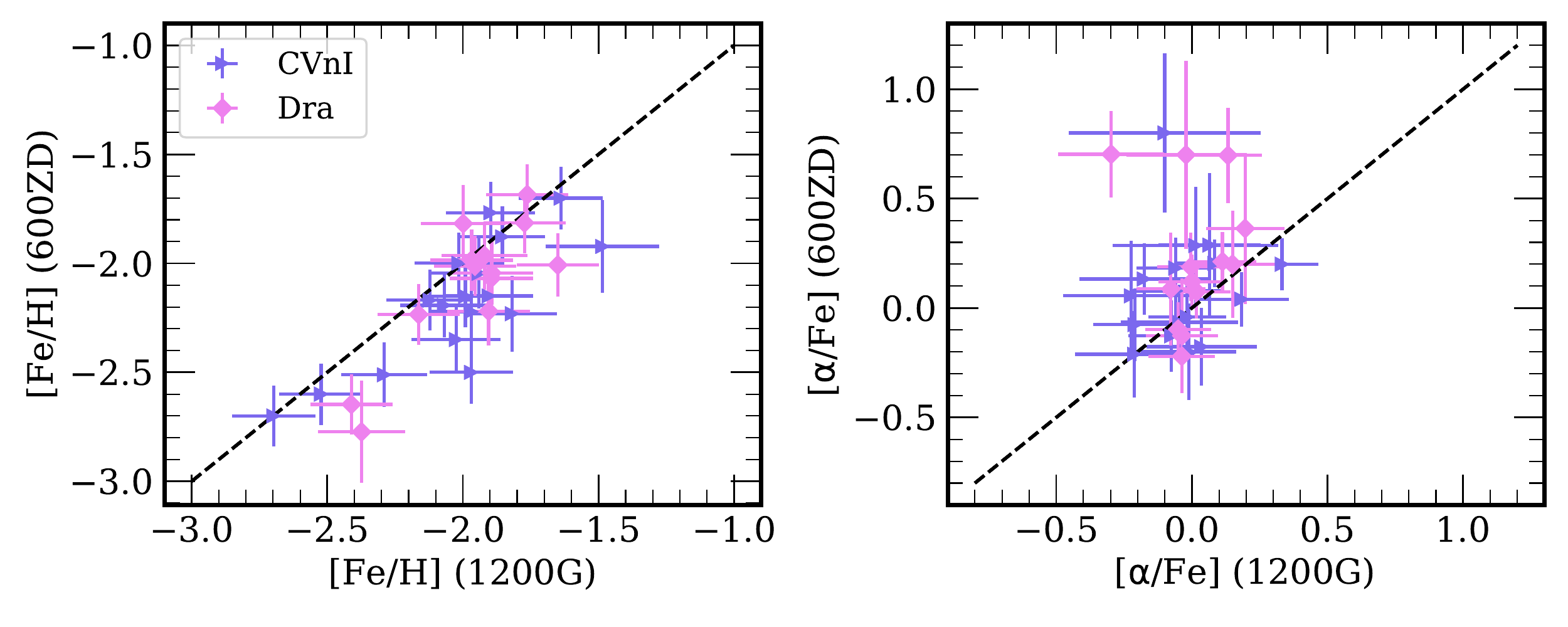}
    \caption{A star-by-star comparison between \feh\ and \alphafe\ for 30 giant stars from a sample of MW dSphs \citep{Kirby2010,Escala2019} using spectra obtained with the 1200G and 600ZD gratings on DEIMOS. The dashed line represents a one-to-one relation. As in the case of the GCs, the measurements are broadly consistent ($\sigma_{\rm [Fe/H]}$ = 1.28, $\sigma_{[\alpha/{\rm Fe}]} = 1.14$) between the two techniques. The 600ZD \feh\ measurements appear to be more metal-poor than the 1200G \feh\ measurements for the dSph stars by $\sim$0.1 dex, although it is unclear if this trend is representative of M31.
    }
    \label{fig:dsphs_appendix}
\end{figure*}

\added{Based on our GC sample, we conclude that the 600ZD and 1200G measurements are broadly consistent (within approximately 1.3$\sigma$). We further illustrate this point using our dSph sample, which spans a larger range of \alphafe\ than the GCs. Figure~\ref{fig:dsphs_appendix} shows a star-by-star comparison between 1200G- and 600ZD-based measurements for the dSphs, where the measurements are broadly consistent within the uncertainties ($\sigma_{\rm [Fe/H]} = 1.28$ and $\sigma_{[\alpha/{\rm Fe}]} = 1.14$). The abundance measurements clearly track one another between the two techniques, most notably in the case of \alphafe\ as compared to the GC sample. In contrast to the GCs, we find an error-weighted mean offset between \feh\ measurements for the dSphs, where $\langle\Delta$[Fe/H]$_{\rm 600ZD-1200G}\rangle$ = $-$0.13 $\pm$ 0.02. It is unclear whether this offset present in the dSph sample is representative of a general offset in \feh\ between 1200G- and 600ZD-based measurements, given that it is not present in our larger GC sample ($N_{\rm dSphs} = 30$ vs. $N_{\rm GCs} = 81$). 
We do not anticipate that this potential offset in \feh\ could significantly alter any of our conclusions as stated in \S~\ref{sec:summary} that rely on a comparison between 1200G- and 600ZD-based abundances, given that the quoted uncertainties in, e.g., $\langle$[Fe/H]$\rangle$, are similar in magnitude.}

\section{The Velocity Dispersion of the Outer Disk of M31}
\label{sec:disk_dispersion}

The dispersion of the disk feature in field D is low (16$^{+3}_{-2}$ \kms; Table~\ref{tab:kinematic_decomp}) compared to expectations based on previous analyses of M31's northeastern disk kinematics at smaller disk radii ($\lesssim$ 20 projected kpc; \citealt{Ibata2005,Dorman2012,Dorman2015}). 
\citet{Collins2011} analyzed slitmasks at similar radii to field D, albeit in M31's southwestern disk.
Although a less stringent velocity cut to identify M31 RGB stars (\S~\ref{sec:membership}) would increase the velocity dispersion of the disk feature in field D, the increase is insufficient to resolve the discrepancy. Assuming that all stars with $v_{\rm{helio}}$ $<$ $-$50 \kms\ in field D are bona fide M31 RGB stars (N$_{\rm star}$ = 73) results in the dispersion of M31's disk increasing to $\sigma_v$ $\sim$25 \kms\, based on estimates from the EM algorithm for fitting Gaussian mixtures to the velocity distribution. 
Entirely removing radial velocity as a criterion for M31 membership (N$_{\rm star}$ = 76), we instead found $\sigma_v$ $\sim$ 40 \kms\ for the disk feature in field D, which is comparable to the values found by \citet{Dorman2012}, \citet{Ibata2005}, and \citet{Collins2011}, although these studies accounted for MW foreground star contamination by various means. Although we used a relatively conservative velocity cut of $v_{\rm{helio}}$ $<$ $-$100 \kms\ to identify M31 RGB field stars in field D, the MW contamination fraction appears to be low in this field based on the absence of an velocity peak at $\sim-$50 \kms\ (Figure~\ref{fig:kinematic_decomp}) corresponding to MW foreground stars.

 

Regardless of the details of sample selection, M31's northeastern disk exhibits intrinsic spatial variation in disk kinematics across its entire radial range, where the velocity dispersion on large scales decreases with increasing disk radius \citep{Dorman2015}. We also expect that the local velocity dispersion of a dynamically cold stellar population will be smaller when computed in individual DEIMOS fields as compared to subregions of the disk with a larger extent in position angle. Measuring the collective velocity dispersion of the disk (e.g., \citealt{Dorman2015}) for studies with wide spatial coverage requires assuming a disk model, which may affect measurements of the velocity dispersion. This likely explains why our measurement is more similar to studies that have averaged velocity dispersion measurements across individual DEIMOS slitmasks in M31's disk \citep{Ibata2005,Collins2011,Dorman2012}. Thus, we conclude that our measured velocity dispersion of $\sim$15$-$20 \kms\ is an accurate representation of the dynamics of the feature \added{that} we \added{have} identified as part of M31's disk. 




\section{Stellar Parameters and Elemental Abundances of Individual M31 RGB Stars}
\label{sec:abund_table}

Here, we present a table of stellar parameters and elemental abundances of individual M31 RGB stars (Table~\ref{tab:abund_catalog}) in the 12 kpc halo (H), 22 kpc GSS (S), 26 kpc disk (D), and 23 kpc halo fields (f130\_2). The table includes data for the 70 M31 RGB stars across all 4 fields  with reliable \feh \ and \alphafe\ measurements (\S~\ref{sec:sample}), in addition to M31 RGB stars that only have \feh \ measurements. Stars with signatures of TiO absorption in their spectra are omitted from the table.

\startlongtable
\begin{deluxetable}{lccccccccccc}
\tablecolumns{12}
\tablewidth{0pc}
\tablecaption{Stellar Parameters and Elemental Abundances of Individual M31 RGB Stars\label{tab:abund_catalog}\tablenotemark{a}}
\tablehead{
        \multicolumn{1}{c}{Object} &
        \multicolumn{2}{c}{Sky Coordinates} & 
        \multicolumn{1}{c}{$v_{\rm{helio}}$} &
        \multicolumn{1}{c}{S/N} &
        \multicolumn{1}{c}{$T_{\rm eff}$} & \multicolumn{1}{c}{$\delta(T_{\rm eff}$)} &\multicolumn{1}{c}{$\log$ $g$} &\multicolumn{1}{c}{\feh} &\multicolumn{1}{c}{$\delta$(\feh)} &\multicolumn{1}{c}{\alphafe} &\multicolumn{1}{c}{$\delta$(\alphafe)}\\
        \multicolumn{1}{c}{ID} &
        \multicolumn{1}{c}{RA} & \multicolumn{1}{c}{Dec} & \multicolumn{1}{c}{(km s$^{-1}$)} & 
        \multicolumn{1}{c}{(\AA\,$^{-1}$)} &
        \multicolumn{1}{c}{(K)} & 
        \multicolumn{1}{c}{(K)} &
        \multicolumn{1}{c}{(dex)} &
        \multicolumn{1}{c}{(dex)} &
        \multicolumn{1}{c}{(dex)} &
        \multicolumn{1}{c}{(dex)} &
        \multicolumn{1}{c}{(dex)}
}
\startdata
\multicolumn{12}{c}{12 kpc Halo Field (H)}\\\hline
1005969 & 00h46m13.39s & +40d40m10.3s & $-$177.4 & 13 & 4472 & 87 & 0.98 & $-$1.06 & 0.14 & \nodata  & \nodata  \\
1007726 & 00h46m24.18s & +40d41m43s & $-$370.0 & 12 & 3875 & 7 & 0.81 & $-$2.92 & 0.27 & \nodata  & \nodata  \\
1007736 & 00h46m12.93s & +40d41m43.2s & $-$391.1 & 6 & 4253 & 12 & 1.27 & $-$2.27 & 0.32 & \nodata  & \nodata  \\
1009083 & 00h46m19.45s & +40d42m45.4s & $-$278.4 & 11 & 3569 & 106 & 0.84 & $-$0.67 & 0.14 & \nodata  & \nodata  \\
1009202 & 00h46m21.6s & +40d42m49.3s & $-$551.7 & 10 & 4100 & 876 & 1.33 & $-$2.10 & 0.23 & \nodata  & \nodata  \\
1009347 & 00h46m25.17s & +40d42m58.7s & $-$276.1 & 15 & 3871 & 4 & 0.81 & $-$1.79 & 0.15 & 0.90 & 0.33 \\
1009577 & 00h46m25.9s & +40d43m07.2s & $-$305.4 & 17 & 3964 & 35 & 0.66 & $-$1.37 & 0.14 & 0.70 & 0.26 \\
1009789 & 00h46m04.69s & +40d43m14.3s & $-$317.0 & 14 & 4029 & 2399 & 0.62 & $-$2.18 & 0.17 & 0.95 & 0.35 \\
\hline
\multicolumn{12}{c}{22 kpc GSS Field (S)} \\\hline
14648 & 00h44m16.87s & +39d48m55.5s & $-$427.7 & 9 & 4800 & 169 & 1.60 & $-$1.22 & 0.16 & \nodata  & \nodata  \\
157934 & 00h44m00.53s & +39d35m51.8s & $-$314.8 & 17 & 4299 & 118 & 0.66 & $-$0.02 & 0.13 & 0.51 & 0.21 \\
169191 & 00h44m02.32s & +39d37m47.1s & $-$472.0 & 10 & 4457 & 165 & 1.39 & $-$0.50 & 0.14 & 0.68 & 0.28 \\
178993 & 00h43m56.03s & +39d39m24.1s & $-$202.1 & 11 & 5257 & 727 & 1.51 & $-$1.35 & 0.20 & \nodata  & \nodata  \\
190457 & 00h44m11.76s & +39d41m14.5s & $-$525.0 & 9 & 3734 & 182 & 1.59 & $-$0.47 & 0.14 & \nodata  & \nodata  \\
2000189 & 00h44m03.97s & +39d37m34s & $-$367.6 & 11 & 3592 & 85 & 0.93 & $-$1.95 & 0.16 & \nodata  & \nodata  \\
2000833 & 00h44m03.16s & +39d38m26.3s & $-$455.2 & 21 & 3903 & 995 & 1.14 & $-$1.49 & 0.15 & 0.06 & 0.37 \\
2001537 & 00h44m04.2s & +39d39m22.2s & $-$513.8 & 11 & 4363 & 8 & 1.30 & $-$1.00 & 0.14 & 0.45 & 0.30 \\
\hline
\multicolumn{12}{c}{26 kpc Disk Field (D)} \\\hline
109460 & 00h49m16.88s & +42d43m53.7s & $-$105.7 & 13 & 3800 & 92 & 0.96 & $-$1.11 & 0.14 & 0.28 & 0.28 \\
16545 & 00h49m03.12s & +42d44m02.5s & $-$105.9 & 32 & 3754 & 149 & 0.74 & $-$2.02 & 0.13 & \nodata  & \nodata  \\
3000373 & 00h48m53.99s & +42d45m23.6s & $-$120.4 & 14 & 3802 & 6 & 0.84 & $-$0.75 & 0.14 & 0.50 & 0.31 \\
3000412 & 00h48m57.21s & +42d45m14.2s & $-$305.4 & 20 & 3729 & 228 & 0.73 & $-$2.71 & 0.16 & 0.10 & 0.47 \\
3000724 & 00h49m02.02s & +42d47m38.1s & $-$152.6 & 15 & 4200 & 88 & 0.82 & $-$0.51 & 0.13 & 1.04 & 0.25 \\
32476 & 00h49m02.3s & +42d45m14s & $-$413.1 & 17 & 4400 & 85 & 1.05 & $-$1.54 & 0.14 & 1.07 & 0.20 \\
57165 & 00h49m06.74s & +42d44m39.4s & $-$127.3 & 13 & 3950 & 64 & 0.79 & $-$0.81 & 0.14 & 0.94 & 0.23 \\
760462 & 00h49m57.17s & +42d41m45.5s & $-$167.3 & 19 & 3798 & 29 & 0.63 & $-$0.89 & 0.13 & 0.57 & 0.20 \\
\hline
\multicolumn{12}{c}{23 kpc Halo Field (f130\_2)} \\\hline
1282152 & 00h50m17.45s & +40d16m31.4s & $-$158.4 & 21 & 4099 & 10 & 0.63 & $-$1.01 & 0.13 & $-$0.04 & 0.26 \\
1282547 & 00h50m11.59s & +40d18m34.9s & $-$361.5 & 11 & 4034 & 9 & 0.90 & $-$1.27 & 0.14 & \nodata  & \nodata  \\
1292468 & 00h49m56.71s & +40d18m19.1s & $-$259.9 & 12 & 3719 & 4 & 0.61 & $-$0.73 & 0.14 & 0.50 & 0.37 \\
1292507 & 00h49m51.47s & +40d18m14.2s & $-$316.5 & 18 & 3911 & 7 & 0.45 & $-$1.76 & 0.14 & 0.60 & 0.34 \\
1292637 & 00h49m46.52s & +40d16m53.1s & $-$161.5 & 6 & 3764 & 8 & 0.87 & $-$2.13 & 0.23 & \nodata  & \nodata  \\
1292654 & 00h49m34.73s & +40d17m32.8s & $-$220.2 & 6 & 3863 & 11 & 1.07 & $-$0.15 & 0.14 & \nodata  & \nodata  \\
1302581 & 00h49m09.27s & +40d15m28.7s & $-$290.2 & 12 & 4603 & 5 & 1.04 & $-$2.57 & 0.21 & 0.40 & 0.50 \\
1302582 & 00h49m27.6s & +40d15m27.4s & $-$161.5 & 11 & 3634 & 6 & 0.72 & $-$2.73 & 0.20 & \nodata  & \nodata  \\
\enddata
\tablenotetext{a}{The errors presented for $T_{\rm eff}$ represent only the random component of the total uncertainty. However, the errors for \feh \ and \alphafe\ include systematic components that account for errors propagated by inaccuracies in $T_{\rm eff}$ \citep{Escala2019}}
\tablecomments{(This table is available in its entirety in machine-readable form.).}
\end{deluxetable}


\end{document}